\documentclass{article}

\newcommand{\typeof}{0}
\usepackage{url}
\usepackage{ifthen}
\newcommand{\condinc}[2]{\ifthenelse{\equal{\typeof}{0}}{#1}{#2}}

\condinc{\usepackage{latex8}}{\usepackage{latex8}}

\usepackage{times}
\usepackage[all]{xy}
\usepackage{proof}
\usepackage{epsfig}
\usepackage{latexsym}
\usepackage{amssymb}
\usepackage{amsmath}
\usepackage{graphicx}
\usepackage{stmaryrd}
\usepackage{subfigure}

\newenvironment{varitemize}
{
\begin{list}{\labelitemi}
{\setlength{\itemsep}{0pt}
 \setlength{\topsep}{0pt}
 \setlength{\parsep}{0pt}
 \setlength{\partopsep}{0pt}
 \setlength{\leftmargin}{15pt}
 \setlength{\rightmargin}{0pt}
 \setlength{\itemindent}{0pt}
 \setlength{\labelsep}{5pt}
 \setlength{\labelwidth}{10pt}}}
{
 \end{list} 
}

\newcounter{numberone}

\newenvironment{varenumerate}
{
\begin{list}{\arabic{numberone}.}
{
  \usecounter{numberone}
  \setlength{\itemsep}{0pt}
  \setlength{\topsep}{0pt}
  \setlength{\parsep}{0pt}
  \setlength{\partopsep}{0pt}
  \setlength{\leftmargin}{15pt}
  \setlength{\rightmargin}{0pt}
  \setlength{\itemindent}{0pt}
  \setlength{\labelsep}{5pt}
  \setlength{\labelwidth}{15pt}
}}
{
\end{list} 
}

\newcounter{numbertwo}

\newenvironment{varvarenumerate}
{
\begin{list}{\roman{numbertwo}.}
{
  \usecounter{numbertwo}
  \setlength{\itemsep}{0pt}
  \setlength{\topsep}{0pt}
  \setlength{\parsep}{0pt}
  \setlength{\partopsep}{0pt}
  \setlength{\leftmargin}{15pt}
  \setlength{\rightmargin}{0pt}
  \setlength{\itemindent}{0pt}
  \setlength{\labelsep}{5pt}
  \setlength{\labelwidth}{15pt}
}}
{
\end{list} 
}

\newcommand{\sublambda}{\Theta}
\newcommand{\sleal}{\Theta_\mathsf{EAL}}
\newcommand{\sllal}{\Theta_\mathsf{LAL}}
\newcommand{\graphs}{\Delta}
\newcommand{\greal}{\Delta_\mathsf{EAL}}
\newcommand{\grlal}{\Delta_\mathsf{LAL}}
\newcommand{\grasr}{\Delta_\mathsf{ASR}}
\newcommand{\trans}{\mathcal{T}}
\newcommand{\trealasr}{\mathcal{T}^\mathsf{EAL}_\mathsf{ASR}}
\newcommand{\trlalasr}{\mathcal{T}^\mathsf{LAL}_\mathsf{ASR}}
\newcommand{\readb}{\mathcal{R}}
\newcommand{\rbasrlm}{\mathcal{R}^\mathsf{ASR}_\Lambda}

\newcommand{\rwgraphs}{\rightarrow_{\graphs}}
\newcommand{\rweal}{\rightarrow_\mathsf{EAL}}
\newcommand{\rwlal}{\rightarrow_\mathsf{LAL}}
\newcommand{\rwasr}{\rightarrow_\mathsf{ASR}}
\newcommand{\nf}[2]{\mathit{NF}(#1)_{#2}}
\newcommand{\labfunc}{\mathcal{F}}
\newcommand{\conseman}[2]{\llbracket #1\rrbracket_{#2}}
\newcommand{\consem}[1]{\llbracket #1\rrbracket}
\newcommand{\linear}{\multimap}

\newcommand{\cs}[2]{#1\sim #2}
\newcommand{\csor}[2]{#1\triangleright #2}
\newcommand{\csora}[2]{#1&\triangleright& #2}
\newcommand{\csea}[2]{#1&\sim &#2}
\newcommand{\pp}[1]{\mathit{pp}(#1)}
\newcommand{\fp}[1]{\mathit{fp}(#1)}
\newcommand{\wpo}[1]{\mathit{wp}(#1)}
\newcommand{\type}[1]{\mathit{ty}(#1)}
\newcommand{\epsi}[1]{\mathit{ep}(#1)}
\newcommand{\spb}{\;|\;}
\newcommand{\head}{\Xi}
\newcommand{\headpn}{\Psi}
\newcommand{\hommap}{\Phi}
\newcommand{\PP}{\mathsf{p}}
\newcommand{\QQ}{\mathsf{q}}
\newcommand{\N}{\mathbb{N}}

\newcommand{\EAL}{\textsf{EAL}}
\newcommand{\EALso}{\ensuremath{\textsf{EAL}_{\forall}}}

\newcommand{\LAL}{\textsf{LAL}}

\newcommand{\LBL}{\textrm{LBL}}
\newcommand{\MLBL}{\textrm{MLBL}}
\newcommand{\tmone}{t}
\newcommand{\tmtwo}{u}
\newcommand{\tmthree}{v}
\newcommand{\tmfour}{s}
\newcommand{\pnone}{N}
\newcommand{\pntwo}{M}
\newcommand{\pnthree}{L}
\newcommand{\sgone}{G}
\newcommand{\sgtwo}{H}

\newcommand{\hsone}{\mathit{X}}
\newcommand{\hstwo}{\mathit{Y}}

\newtheorem{definition}{Definition}

\newtheorem{lemma}{Lemma}
\newtheorem{proposition}{Proposition}
\newtheorem{theorem}{Theorem}

\newenvironment{proof}{\begin{trivlist}
       \item[\hskip \labelsep {\bfseries Proof.}]}{\hfill$\Box$\end{trivlist}}

\title{Light Logics and Optimal Reduction:\\ Completeness and Complexity}

\author{Patrick Baillot \and Paolo Coppola \and Ugo Dal Lago}

\author{
Patrick Baillot\thanks{Partially supported by projects NO-CoST (ANR, JC05\_43380), CRISS (ACI).}\\
{\it LIPN, CNRS \& Universit\'e Paris Nord}\\
\texttt{pb@lipn.univ-paris13.fr}\\
\and
Paolo Coppola\thanks{Partially supported by a visiting professorship of Univ. Paris Nord.}\\
{\it Universit\`a di Udine}\\
\texttt{coppola@uniud.it}\\
\and
Ugo Dal Lago\thanks{Work partially carried out while the author was postdoctoral fellow of Univ. Paris Nord and
of Univ. Denis Diderot, Paris 7.}\\
{\it Universit\`a di Bologna}\\ 
\texttt{dallago@cs.unibo.it}\\
}

\begin{document}
\setlength{\parskip}{0pt}
\maketitle
\begin{abstract}

 Typing of lambda-terms in Elementary and Light Affine Logic (\EAL\ , \LAL\,  resp.) has
been studied for two different reasons: on the one hand the evaluation
of typed terms using \LAL\  (\EAL\, resp.) proof-nets admits a guaranteed polynomial 
(elementary, resp.) bound;
on the other hand these terms can also be evaluated by optimal reduction using
the abstract
version of Lamping's algorithm. The first reduction is global while the second
one is local and asynchronous. We prove that for \LAL\  (\EAL\, resp.) typed terms,
Lamping's abstract algorithm also admits a polynomial (elementary, resp.) bound.
We also show its soundness and completeness (for \EAL\  and \LAL\  with type
fixpoints), by using a simple geometry of interaction model (context semantics).

 \end{abstract}
\section{Introduction}
\paragraph{Background and Motivations.}
Light logics such as Light Affine Logic (\LAL) and Elementary Affine
Logic (\EAL) have been introduced in \cite{Girard98,Asperti98} and
then studied as type systems for lambda-calculus
\cite{AspertiRoversi02,coppola06tocl,coppola05fi,ABT06,
baillot04lics,coppola05tlca} and semantically (\textit{e.g.} in
\cite{LaurentTortora06,MurawskiOng00,DalLagoHofmann05}). Their analysis has been
motivated by two distinct features:
\begin{varenumerate}
\item \textit{Complexity-certified reduction}: using the syntax of 
\textit{proof-nets}, the logics \LAL\  (\EAL, respectively)  ensure that the program terminates
with a polynomial (elementary, respectively) time bound;
\item \textit{Simple optimal reduction}: for lambda-terms typed in these systems
one can use the abstract version of Lamping's algorithm~\cite{Lamping90}, without 
the so-called \textit{oracle}, so
plain graph rewriting with local rules.
\end{varenumerate}
 However each of these approaches has its drawbacks:
\begin{varitemize}
\item 
  Proof-net reduction  is global, requires the handling of \textit{boxes}, and thereby
enforces a form of synchronicity which is awkward to  implement;
\item 
  Optimal reduction is local, asynchronous --  and elegant, but \dots does not offer
  any guaranteed complexity bound. 
\end{varitemize}
Actually, even the fact that the abstract Lamping algorithm is \textit{correct} for
terms of \EAL\ and \LAL\ is not completely straightforward: it was pointed out 
since~\cite{Asperti98} and is part of the folklore,
but it seems no explicit direct proof of this statement is available
in the literature, although the general techniques of \cite{guerrini03tcs}
could be applied to this restricted setting.

 The goal of this paper is therefore to bring together the best of these two worlds:
we prove that terms typable in \LAL\ (\EAL, respectively) can be reduced by the Lamping
abstract algorithm with a certified polynomial (elementary, respectively) time
bound. Moreover a type derivation in \LAL\ or \EAL\ carries two kinds of information: 
the sharing information and the boxing information. We actually show here that the boxing information
is \textit{not} needed to perform Lamping's abstract algorithm. Some systems
like \textsf{DLAL} \cite{baillot04lics} or restrictions of \EAL\ \cite{coppola06tocl} do not
use sharing: in that case knowing that the term is typeable (without knowing the type)
is sufficient to apply the abstract algorithm. 

 Actually the bounds of light logics can also be obtained without the proof-net machinery,
in plain lambda-calculus, if one considers fragments of the type systems, possibly with
restricted (lazy) reduction~\cite{baillot04lics,coppola05tlca}. However this is still
a global form of reduction ($\beta$-reduction). Here we aim to handle the full type systems
and to switch to a local reduction, which is motivating for concrete implementations and in 
particular for distributed evaluation~\cite{PediciniQuaglia00}.

\paragraph{Optimal Reduction and Light Logics.}
%
%
The fact that \EAL\ typable terms can be reduced
with Lamping's abstract algorithm
is quite remarkable, since it is known that the bookkeeping
needed for the oracle causes inefficiencies in optimal reduction~\cite{Lawall96icfp}.


On the other hand, proof-net reduction in these systems
is performed with guaranteed complexity bound,
one might think that the preservation of bounds when switching
from proof-net reduction to optimal reduction is a consequence of
optimality itself. However  this is actually not true: the optimality 
concerns the number of parallel beta-steps, which is not 
directly related to the normalisation time \cite{AspertiMairson01,AspertiCoppolaMartini04}.
For an in-depth study of optimal reduction one can consult \cite{AspertiGuerrini98}.

 Moreover, 
techniques used when analyzing proof-net (or lambda-term) reduction time
cannot be directly applied here. In particular, the level-by-level reduction
strategy (see \cite{Girard98,AspertiRoversi02}) has no counterpart in
 the framework of sharing graphs, where
copying is done incrementally.

\paragraph{Contributions.}
Our main results  are:
\begin{varitemize}
\item
  We define a general class of admissible translations from light logics type
  derivations to sharing graphs, subsuming the ones proposed before.
\item
  For each admissible translation, we show that graph reduction is
  sound and complete with respect to lambda-reduction.
\item
  Moreover, we show that graph reduction can be performed in bounded
  time, where the bound is
of the same complexity order as the one we have on the underlying logical system. 
\end{varitemize}
Moreover we believe that the main technique used to prove the complexity bounds
(Section \ref{sect:complexity}),
based on the definition of \textit{weights} for sharing graphs (or interaction nets, 
\cite{Lafont97})
following \cite{DalLagoArxiv2005d}
is of its own interest and could presumably be applied to other problems.

\condinc{}{A full version with complete proofs is available \cite{BCDL07}.}
\section{Soundness and Completeness in the General Case}
Before introducing the specific logical systems we are interested in, we define
the notions of soundness and completeness for abstract systems of
graph reduction.
Throughout the paper, $\Lambda$ is the set of pure, untyped, lambda
terms.  If $A$ is a set and $\rightarrow$ is a binary relation on $A$,
the set of normal forms in $A$ (with respect to $\rightarrow$) will be
denoted $\nf{A}{\rightarrow}$.

\begin{definition}[Graph Rewriting Systems]
A $\sublambda$-graph rewriting system is a quintuple $(\sublambda,\graphs,\rwgraphs,\trans,\readb)$ where:
\begin{varitemize}
\item
  $\sublambda\subseteq\Lambda$ is a set of lambda-terms to which the technique can be applied.
\item
  $\graphs$ is a set of \emph{graphs}.
\item
  $\rwgraphs$ is a rewriting relation on $\graphs$.
\item
  $\trans$ is a total binary relation from $\sublambda$ to $\graphs$, called the \emph{initial translation}.
\item
  $\readb$ is a function from $\graphs$ to $\Lambda$, called the \emph{readback}.
\end{varitemize}
\end{definition}
Note that $\trans$ is a relation and not a mere function, since we want to allow
several possible translations of a term (this is related to the fact
that we will allow the possibility to decorate a given lambda-term
as several different proof-nets).  

\begin{definition}[Soundness]
We say that the $\sublambda$-graph rewriting system $(\sublambda,\graphs,\rwgraphs,\trans,\readb)$
is \emph{sound} with respect to a reduction relation $\rightarrow$ on $\Lambda$ 
iff for every term $\tmone\in\sublambda$, if $\sgone\in\trans(\tmone)$ and $\sgone$ 
reduces to normal form $\sgtwo$ (in $\rwgraphs$) then $\tmone$ reduces to normal form $\tmtwo$
(in $\rightarrow$)  and $\readb(\sgtwo)=\tmtwo$:
\condinc{
\begin{displaymath}
\xymatrix{\tmone \ar @{-->} [r]^>{*} \ar[d]^{\trans} & \tmtwo\\
\sgone\ar[r]^>{*}_>{\graphs} & \sgtwo\ar[u]_{\mathcal{R}}}
\end{displaymath}}{}
\end{definition}
Soundness of a $\sublambda$-graph rewriting system implies that if we start with a term
$t$ in $\sublambda$, translate it into a graph, reduce the graph and finally read-back
a term $u$, then $u$ is the normal form of $t$. This does not mean the $\sublambda$-graph
rewriting system will necessarily do its job: to be sure about that, we need 
completeness:

\begin{definition}[Completeness]
We say that the $\Theta$-graph rewriting system $(\sublambda,\graphs,\rwgraphs,\trans,\readb)$
is \emph{complete} with respect to a reduction relation $\rightarrow$ on $\Lambda$ 
iff for every term $\tmone\in\sublambda$ if $\tmone$ reduces to normal form
$\tmtwo$, then any $\sgone\in\trans(\tmone)$ reduces to normal form 
$\sgtwo$, where $\readb(\sgtwo)=\tmtwo$.
\condinc{
\begin{displaymath}
\xymatrix{\tmone \ar[r]^>{*}\ar[d]^{\trans} & \tmtwo\\
\sgone\ar @{-->} [r]^>{*}_>{\graphs} & \sgtwo\ar[u]_{\readb}}
\end{displaymath}}{
\begin{figure}
\vspace{-5mm}
\begin{center}
  $$
  \begin{array}{cc}
      \xymatrix{\tmone \ar @{-->} [r]^>{*} \ar[d]^{\trans} & \tmtwo\\
        \sgone\ar[r]^>{*}_>{\graphs} & \sgtwo\ar[u]_{\mathcal{R}}}
    &
      \xymatrix{\tmone \ar[r]^>{*}\ar[d]^{\trans} & \tmtwo\\
        \sgone\ar @{-->} [r]^>{*}_>{\graphs} & \sgtwo\ar[u]_{\readb}}
  \end{array}
  $$
\end{center}
\vspace{-5mm}
\caption{Soundness and Completeness}
\vspace{-5mm}
\end{figure}
}
\end{definition}

\section{Type Assignment Systems and Proof-Nets}

%
Formulae of (intuitionistic) \emph{elementary affine logic} (\EAL\ for short)
 are generated by the following
productions:
$$
A::=\alpha\spb A\linear A\spb !A\spb\forall\alpha.A\spb\mu\alpha.A
$$
where $\alpha$ ranges over a countable set of \emph{atoms}. Recall that $!$ is called
an \textit{exponential connective} or \emph{modality}.
 
Here we are considering in fact \EAL\ with type fixpoints (recursive types)
but this does not modify its normalisation properties~\cite{DalLagoBaillot2006}. 
Most references in the literature 
deal with the second order fragment \EALso, which does not include type fixpoints.

\EAL\ can be seen as a type system for terms
in $\Lambda$:  Figure~\ref{figure:ealsc} presents type assignment by means
of sequent calculus, which is tedious for typing but convenient
for studying the dynamics. Other presentations of typing 
can be found in the literature~\cite{coppola06tocl,coppola05fi,BaillotTerui05}. Note that
sharing is allowed, for instance by using rules $X$ and $U$. 
 $\sleal$ denotes
the set of lambda terms which are typable in elementary affine logic.

\begin{figure}
\begin{center}
{\footnotesize
\begin{minipage}{.46\textwidth}
\centering\textbf{Axiom, Cut and Structural Rules}
$$
\begin{array}{ccc}
\infer[A]{x:A\vdash x:A}{} 
&&
\infer[U]
{\Gamma,\Delta\vdash \tmtwo\{\tmone/x\}:B} 
{\Gamma\vdash \tmone:A & \Delta,x:A\vdash \tmtwo:B}
\end{array}
$$
$$
\begin{array}{ccc}
\infer[W]
{\Gamma,x:A\vdash \tmone:B}
{\Gamma\vdash \tmone:B}
&&
\infer[X]
{\Gamma,z:!A\vdash \tmone\{z/x,z/y\}:B}
{\Gamma,x:!A,y:!A\vdash \tmone:B}
\end{array}
$$
\centering\textbf{Multiplicative Logical Rules}
$$
\begin{array}{cc}
\infer[R_\multimap]
{\Gamma\vdash \lambda x.\tmone:A\multimap B}
{\Gamma,x:A\vdash \tmone:B} 
&
\infer[L_\multimap]
{\Gamma,\Delta,y:A\multimap B\vdash \tmtwo\{y\tmone/x\}:C} 
{\Gamma\vdash \tmone:A & \Delta,x:B\vdash \tmtwo:C}
\end{array}
$$
\centering\textbf{Exponential Rules}
$$
\infer[P_{!}]
{!\Gamma\vdash\tmone:!A}
{\Gamma\vdash\tmone:A}
$$
\centering\textbf{Second Order Rules}
$$
\begin{array}{ccc}
\infer[R_{\forall}]
{\Gamma\vdash \tmone:\forall\alpha.A}
{\Gamma\vdash \tmone:A & \alpha\not\in\mathit{FV}(\Gamma)}
&&
\infer[L_{\forall}]
{\Gamma,x:\forall\alpha.A\vdash \tmone:C}
{\Gamma,x:A\{B/\alpha\}\vdash \tmone:C}
\end{array}
$$
\centering\textbf{Least Fixpoint Rules}
$$
\begin{array}{ccc}
\infer[R_{\mu}]
{\Gamma\vdash \tmone:\mu\alpha.A}
{\Gamma\vdash \tmone:A\{\mu\alpha.A/\alpha\}}
&&
\infer[L_{\mu}]
{\Gamma,x:\mu\alpha.A\vdash \tmone:B}
{\Gamma,x:A\{\mu\alpha.A/\alpha\}\vdash \tmone:B}
\end{array}
$$
\end{minipage}}
\caption{A sequent calculus for elementary linear logic with second order and fixpoints.}
\label{figure:ealsc}
\end{center}
\vspace{-10mm}
\end{figure}
\begin{figure*}
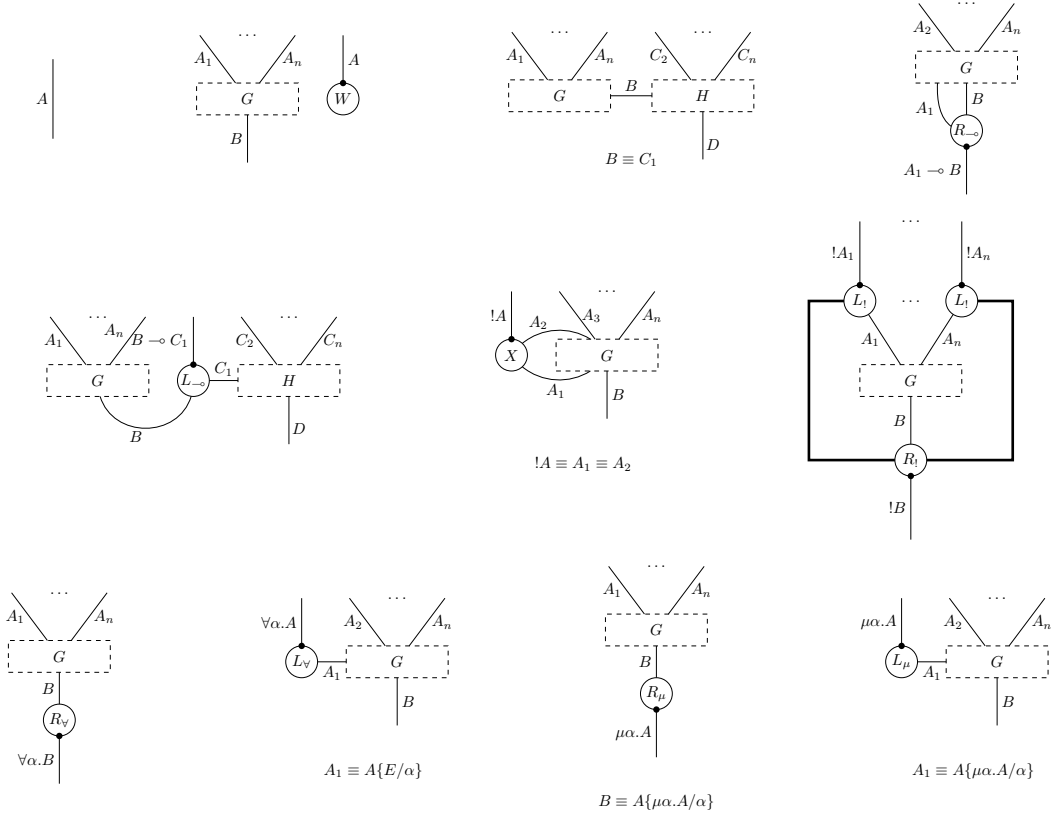
 
\begin{center}
  \begin{minipage}[c]{7.2pt}
    \centering\scalebox{0.6}{\epsfbox{figure.8}}
  \end{minipage}\hspace{50pt}
  \begin{minipage}[c]{63pt}
    \centering\scalebox{0.6}{\epsfbox{figure.10}}
  \end{minipage} \hspace{50pt}
  \begin{minipage}[c]{95.4pt}
    \centering\scalebox{0.6}{\epsfbox{figure.5}}
  \end{minipage} \hspace{50pt}
  \begin{minipage}[c]{45pt}
    \centering\scalebox{0.6}{\epsfbox{figure.3}}
  \end{minipage} \\ \vspace{8pt}
  \begin{minipage}[c]{113.4pt}
    \centering\scalebox{0.6}{\epsfbox{figure.4}}
  \end{minipage} \hspace{50pt}
  \begin{minipage}[c]{65.4pt}
    \centering\scalebox{0.6}{\epsfbox{figure.25}}
  \end{minipage} \hspace{50pt}
  \begin{minipage}[c]{78pt}
    \centering\scalebox{0.6}{\epsfbox{figure.7}}
  \end{minipage} \\ \vspace{8pt}
  \begin{minipage}[c]{41.4pt}
    \centering\scalebox{0.6}{\epsfbox{figure.50}}
  \end{minipage} \hspace{50pt}
  \begin{minipage}[c]{72.6pt}
    \centering\scalebox{0.6}{\epsfbox{figure.51}}
  \end{minipage} \hspace{50pt}
  \begin{minipage}[c]{44.4pt}
    \centering\scalebox{0.6}{\epsfbox{figure.52}}
  \end{minipage} \hspace{50pt}
  \begin{minipage}[c]{72.6pt}
    \centering\scalebox{0.6}{\epsfbox{figure.53}}
  \end{minipage}
\end{center}
\caption{Proof-nets for elementary affine logic.}
\label{figure:ealpn}
\end{figure*}

Elementary affine logic proofs can be formulated as a system of
(intuitionistic) proof-nets $\greal$, defined inductively on
Figure~\ref{figure:ealpn}. Node $X$ is called a \textit{contraction} node.
The \emph{principal edge} of a node $v$
is the edge incident to $v$ through its \emph{principal
port} (indicated with a  $\bullet$).
A \emph{cut} is an edge $e=\{v,w\}$ which is principal for both $v$ and $w$.

 \EAL\ proof-nets can be
endowed with a rewriting relation $\rweal$ \condinc{(see
Figure~\ref{figure:ealpnrw})}{ that we do not report here for lack of space 
(see~\cite{DalLagoArxiv2005d})}. The important case of  $\rweal$ is when an 
$X$ node meets a $R_{!}$ node, corresponding to a box: in this case
the box is duplicated and the doors $L_{!}$ of the two copies are linked to  
$X$ nodes (\textit{contraction normalisation step}).
 
If  $v$ (resp. $e$) is a node (resp. edge) of a proof-net, $\partial(v)$ (resp. $\partial(e)$)
denotes its \emph{box-depth} (\textit{level}). If $\sgone\in\greal$ is a proof-net, its depth
$\partial(\sgone)$ is the maximal depth of its edges. 
 The \textit{stratification property}
of \EAL\ states that the depth $\partial(e)$ of an edge
does not change through $\rweal$.

\condinc{
\begin{figure}
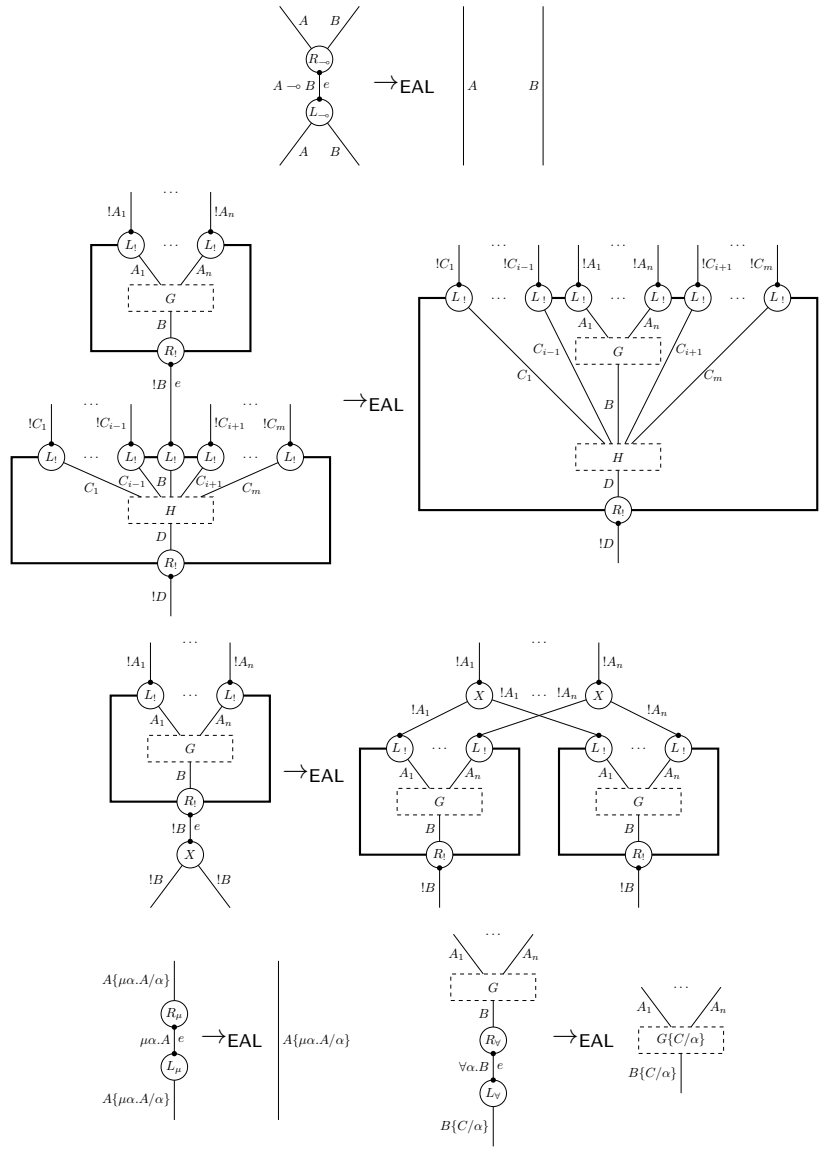

\begin{center}
  \begin{minipage}[c]{34pt}
    \centering\scalebox{0.5}{\epsfbox{figure.14}}
  \end{minipage}
  \begin{minipage}[c]{28pt}
    \centering $\rweal$
  \end{minipage}
  \begin{minipage}[c]{41.5pt}
    \centering\scalebox{0.5}{\epsfbox{figure.15}}
  \end{minipage}\\ \vspace{8pt}
    \begin{minipage}[c]{121pt}
    \centering\scalebox{0.5}{\epsfbox{figure.23}}
  \end{minipage}
  \begin{minipage}[c]{28pt}
    \centering $\rweal$
  \end{minipage}
  \begin{minipage}[c]{152pt}
    \centering\scalebox{0.5}{\epsfbox{figure.24}}
  \end{minipage}\\ \vspace{8pt}
  \begin{minipage}[c]{61pt}
    \centering\scalebox{0.5}{\epsfbox{figure.44}}
  \end{minipage}
  \begin{minipage}[c]{28pt}
    \centering $\rweal$
  \end{minipage}
  \begin{minipage}[c]{137pt}
    \centering\scalebox{0.5}{\epsfbox{figure.45}}
  \end{minipage}\\ \vspace{8pt}
  \begin{minipage}[c]{33pt}
    \centering\scalebox{0.5}{\epsfbox{figure.54}}
  \end{minipage}
  \begin{minipage}[c]{28pt}
    \centering $\rweal$
  \end{minipage}
  \begin{minipage}[c]{29pt}
    \centering\scalebox{0.5}{\epsfbox{figure.55}}
  \end{minipage}
  \hspace{28pt}
  \begin{minipage}[c]{38pt}
    \centering\scalebox{0.5}{\epsfbox{figure.56}}
  \end{minipage}
  \begin{minipage}[c]{28pt}
    \centering $\rweal$
  \end{minipage}
  \begin{minipage}[c]{38pt}
    \centering\scalebox{0.5}{\epsfbox{figure.57}}
  \end{minipage}
\caption{Rewriting rules for elementary affine logic proof-nets.}
\label{figure:ealpnrw}
\end{center}
\end{figure}}{}

\paragraph{Light Affine Logic.} \LAL\ can be obtained from \EAL\ by adopting a 
stricter exponential discipline: one restricts the rule $P_{!}$
of \EAL\ to the case where $\Gamma$ contains at most one
formula, but also adds a new connective $\S$ with
rule $P_{\S}$ (see Figure~\ref{figure:lalsc}). The connective $\S$ is
a weak form of $!$, that does not allow for contraction
(rule $X$). 

 There is a translation $(.)^e$ from \LAL\ to \EAL\ formulae
obtained by replacing $\S$ with $!$. It extends to a translation
on proofs.
Therefore the set  $\sllal$ of lambda-terms typable in \LAL\ is
included in $\sleal$.

 The proofs-nets of \LAL\ are defined as those of \EAL\, but with two new nodes
$L_{\S}$ and $R_{\S}$ and conditions on boxes: a box
with $R_{!}$ main door ($!$-box) can have
at most one $L_{!}$ door; a box with 
  $R_{\S}$ main door ($\S$-box) can have any number of
 $L_{\S}$ and $L_{!}$ doors. A rewriting relation $\rwlal$
is defined on these proof-nets~\cite{DalLagoArxiv2005d}.
This reduction does not cause any duplication of
a $\S$-box.

 The translation $(.)^e$ can be extended naturally
to a translation from \LAL\ to \EAL\ proof-nets,
and it maps  $\rwlal$ to $\rweal$. Therefore the set
of \LAL\ proof-nets can be seen as a subset of $\greal$. Hence properties
of \EAL\ proof-nets will be valid in particular for \LAL\
proof-nets and we will state them only for \EAL\ (except for
complexity issues in Section \ref{sect:complexity}).

\begin{figure}
{\footnotesize
\begin{center}
 \begin{minipage}{.46\textwidth}
$$
\begin{array}{ccccc}
\infer[P_!^1]
{\vdash\tmone:!A}
{\vdash\tmone:A}
&&
\infer[P_!^2]
{x:!A\vdash\tmone:!A}
{x:A\vdash\tmone:A}
&&
\infer[P_{\S}]
{\S\Gamma,!\Delta\vdash\tmone:\S A}
{\Gamma,\Delta\vdash\tmone:A}
\end{array}
$$
\end{minipage}
\caption{Exponential rules of light affine logic with second order and fixpoints.}
\label{figure:lalsc}
\end{center}}
\vspace{-10mm}
\end{figure}

\begin{figure*}
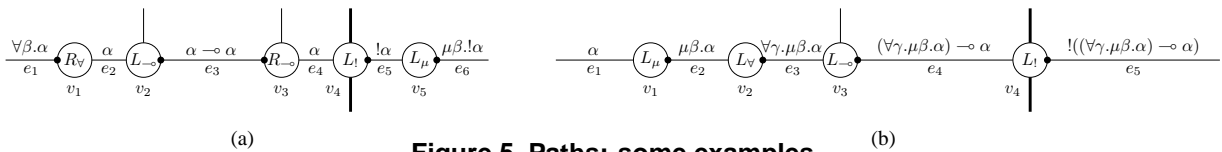

  \begin{center}
    \subfigure[]
    {
      \scalebox{0.65}{\epsfbox{figure.79}}
    }
    \hspace{.5cm}
    \subfigure[]
    {
      \scalebox{0.65}{\epsfbox{figure.80}}
    }
  \end{center}
\vspace{-10mm}
  \caption{Paths: some examples}
  \label{figure:expaths}
\vspace{-5mm}
\end{figure*}

\paragraph{Paths.}
A \emph{direct path} is a sequence of edges $e_1,\ldots,e_n$ such
that the following conditions hold:
\begin{varitemize}
\item
  For every $1\leq i< n$, $e_i$ and $e_{i+1}$ have a vertex $v_i$
  in common;
\item
  For every $1\leq i< n$, $e_i\neq e_{i+1}$ and either $e_i$ or $e_{i+1}$
  is principal for $v_i$.
\end{varitemize}
An example of a direct path is reported in Figure~\ref{figure:expaths}(a).
We say that a direct path $e_1,\ldots,e_n$ with $n\geq 2$ 
\emph{starts} at $v$ iff $e_1=\{v,w\}$ is principal for
$v$ and there is $z$ with $e_2=\{w,z\}$.
A direct path $e_1,\ldots,e_n$ is \emph{simple} iff 
for every $1\leq i<n$, the edge $e_{i+1}$ is principal for
$v_i$. The direct path in Figure~\ref{figure:expaths}(b)
is simple, while the one in Figure~\ref{figure:expaths}(a)
is not. A direct path is \emph{maximal} iff it is not part
of any longer direct path. 
Two edges $e,g$ are \emph{non-consecutive} iff there cannot be any
direct path in the form $e,g$ (or, equivalently, in the form
$g,e$).
 A box $b$ in a 
proof-net $\pnone$ is \emph{special} iff any direct path starting 
from one of its premises is simple.
\begin{lemma}\label{lemma:nonsimple}
Any non-simple direct path $e_1,\ldots,e_n$ starting at any node $v$ contains
a cut $e_i$ such that $\partial(e_i)\leq\partial(e_1)$.
\end{lemma}
\condinc{
\begin{proof}
As a preliminary fact, take notice that for any
simple direct path $e_1,\ldots,e_n$, it holds that $\partial(e_n)\leq\partial(e_1)$.
Indeed, you can enter a box only through a principal port.
We can prove the lemma by induction on $n$:
\begin{varitemize}
\item
  If $n=1$, then the path is simple.
\item
  Then, observe that any non-simple, direct path $e_1,e_2$ starting in $v$ 
  contains a cut, namely $e_1$. Indeed, by definition $e_1$ is principal for $v$ and,
  since the path is non-simple, $e_1$ is principal for $v_1$, too.
\item
  Let $n\geq 3$ and $e_1,\ldots,e_n$ be a non-simple direct path starting from $v$.
  If $e_1,\ldots,e_{n-1}$ is non-simple, then by inductive hypothesis, it 
  contains a cut. If $e_1,\ldots,e_{n-1}$ is simple and $e_1,\ldots,e_n$
  is not simple, then $e_{n-1}$ is principal for $v_{n-2}$ and $v_{n-1}$.
  As a consequence, $e_{n-1}$ is a cut. Moreover, $\partial(e_{n-1})\leq
  \partial(e_1)$.
\end{varitemize}
This concludes the proof.
\end{proof}}{}

\paragraph{Strategies.}
There are two reduction strategies for
proof-nets in $\greal$ (or $\grlal$) that are of particular interest for our
purposes:
\begin{varitemize}
  \item
    The \emph{level-by-level strategy}, \LBL. A cut at level $n+1$ cannot be
    reduced if there are cuts at level $n$.
  \item
    The \emph{modified level-by-level strategy}, \MLBL. It is the level-by-level one 
    with an additional constraint: whenever we copy a box $b$, $b$ must be a special
    box.
\end{varitemize}
\condinc{
Notice that \MLBL\ is a reduction strategy due to Lemma~\ref{lemma:nonsimple}.
Indeed, if a box $b$ is involved in a cut $e$ but is not special, then we can
find another cut $g$ such that $\partial(g)\leq\partial(e)$. But $g$
could be itself an exponential cut involving a non-special box. This sequence
of non-special boxes must however be finite, because otherwise we would have a cycle
that cannot appear in any proof-net (correctness criterion).}
{The fact that \MLBL\ is a reduction strategy is a consequence of Lemma~\ref{lemma:nonsimple}
and of the fact that proof-nets satisfy the correctness criterion~\cite{Girard98}}.

\paragraph{Complexity Bounds.} We can now recall the main
results on \EAL\ and \LAL\ :
\begin{theorem}[Girard~\cite{Girard98}]\label{th:pnbound}
For every natural number $n$, there is a polynomial (respectively, elementary function) 
$e_n:\N\rightarrow\N$ such that for every  proof-net
$\pnone$ of \LAL\ (respectively, \EAL\ )
if $\pnone\rightarrow^n \pntwo$ in the MLBL strategy, then
$n\leq e_{\partial(\pnone)}(|\pnone|)$ and
$|\pntwo|\leq e_{\partial(\pnone)}(|\pnone|)$.
\end{theorem}

Recall that  binary lists can be
represented in \LAL\ with the type:
$W= \forall \alpha. !(\alpha \multimap \alpha) \multimap  !(\alpha \multimap \alpha)
\multimap \S (\alpha \multimap \alpha)$.
This way, a term of \LAL\ type  $t:W \multimap \S^k W$
can be converted to a proof-net, and its application to 
a list evaluated in polynomial time using $\rwlal$. However this
is still a global evaluation procedure and we want to replace it
by optimal reduction.

\section{Lamping's Abstract Algorithm}\label{sect:Lamping}

\begin{figure}
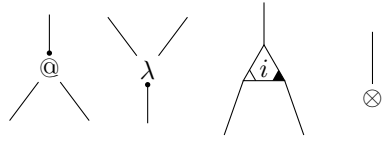

\begin{center}
  \begin{minipage}[c]{34pt}
    \centering\scalebox{1}{\epsfbox{figure.70}}
  \end{minipage}
  \begin{minipage}[c]{41.5pt}
    \centering\scalebox{1}{\epsfbox{figure.71}}
  \end{minipage}
  \begin{minipage}[c]{41.5pt}
    \centering\scalebox{1}{\epsfbox{figure.72}}
  \end{minipage}
  \begin{minipage}[c]{41.5pt}
    \centering\scalebox{1}{\epsfbox{figure.74}}
  \end{minipage}\\ \vspace{8pt}
\caption{Sharing Nodes.}
\label{figure:asr}
\end{center}
\vspace{-10mm}
\end{figure}

\begin{figure*}
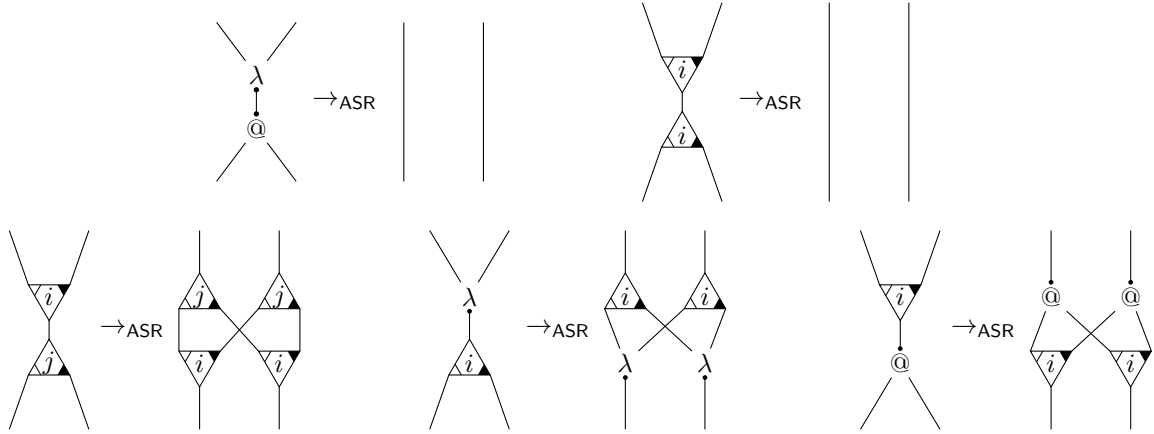

\begin{center}
  \begin{minipage}[c]{34pt}
    \centering\scalebox{1}{\epsfbox{figure.60}}
  \end{minipage}
  \begin{minipage}[c]{28pt}
    \centering $\rwasr$
  \end{minipage}
  \begin{minipage}[c]{41.5pt}
    \centering\scalebox{1}{\epsfbox{figure.61}}
  \end{minipage}\hspace{50pt}
  \begin{minipage}[c]{34pt}
    \centering\scalebox{1}{\epsfbox{figure.62}}
  \end{minipage}
  \begin{minipage}[c]{28pt}
    \centering $\rwasr$
  \end{minipage}
  \begin{minipage}[c]{41.5pt}
    \centering\scalebox{1}{\epsfbox{figure.63}}
  \end{minipage}\\ \vspace{8pt}
  \begin{minipage}[c]{34pt}
    \centering\scalebox{1}{\epsfbox{figure.64}}
  \end{minipage}
  \begin{minipage}[c]{28pt}
    \centering $\rwasr$
  \end{minipage}
  \begin{minipage}[c]{41.5pt}
    \centering\scalebox{1}{\epsfbox{figure.65}}
  \end{minipage}\hspace{50pt}
  \begin{minipage}[c]{34pt}
    \centering\scalebox{1}{\epsfbox{figure.66}}
  \end{minipage}
  \begin{minipage}[c]{28pt}
    \centering $\rwasr$
  \end{minipage}
  \begin{minipage}[c]{41.5pt}
    \centering\scalebox{1}{\epsfbox{figure.67}}
  \end{minipage}\hspace{50pt}
  \begin{minipage}[c]{34pt}
    \centering\scalebox{1}{\epsfbox{figure.68}}
  \end{minipage}
  \begin{minipage}[c]{28pt}
    \centering $\rwasr$
  \end{minipage}
  \begin{minipage}[c]{41.5pt}
    \centering\scalebox{1}{\epsfbox{figure.69}}
  \end{minipage}\\ \vspace{8pt}
\caption{Rewriting rules for sharing graphs.}
\label{figure:asrrw}
\end{center}
\vspace{-10mm}
\end{figure*}

Now we turn to the local reduction procedure.
\condinc{The set of \emph{abstract sharing graphs} $\grasr$ is given by the nodes of 
Figure~\ref{figure:asr}: the 3rd node is called
\textit{fan} and is given together with an
integer \textit{index} $i$.}
{\emph{Abstract sharing graphs} in $\grasr$ are defined by the nodes of 
Figure~\ref{figure:asr}: the 3rd node is called
\textit{fan} and is given together with an
integer \textit{index} $i$.
}
A rewriting relation $\rwasr$ on $\grasr$ is defined on Figure~\ref{figure:asrrw}.
Notice that we omit the garbage collection rules.
This omission is anyway harmless: the readback procedure \condinc{(Section \ref{subsect:readbackprocedure})}{} 
is not affected by the presence of garbage and the complexity
of garbage collection is linear in the size of the graph.

\condinc{If $\sgone$ is a sharing graph, $\fp{\sgone}$ is the set of its \textit{free ports}
(dangling edges), while
$\wpo{\sgone}$ is the set of edges which are incident to $\otimes$-nodes.
If $u$ is a node of $\sgone$, then $\pp{u}$ is the principal port
of $u$.}
{If $\sgone$ is a sharing graph, $\fp{\sgone}$ is the set of its \textit{free ports}
(dangling edges) and
$\wpo{\sgone}$ that of edges incident to $\otimes$-nodes.
If $u$ is a node, then $\pp{u}$ is its principal port. Principal ports for lambda and application nodes are marked with a point in Figure~\ref{figure:asr}. The principal port of the fan is the unmarked one.
}

\condinc{
To translate proof-nets into sharing graphs we will turn contraction nodes into fans. However
we need to choose the indices for the fans. For that, any  proof-net $\pnone$ is given
together with a \emph{labelling function} $\labfunc$ from the set
of its contraction nodes to natural numbers.  The translation $\trealasr$ 
from proof-nets to abstract sharing graphs will be defined up to such labelling functions. 
$\trealasr(\pnone,\labfunc)$ is the graph $\sgone \in \grasr$ obtained in
the following way:
}
{To translate proof-nets into sharing graphs we need to set the indices for the fans.
 For that, any  proof-net $\pnone$ is given
together with a \emph{labelling function} $\labfunc$ from the set
of its contraction nodes to natural numbers. We require the  function $\labfunc$ for the proof-net $\pnone$ 
to be \textit{compatible with depths}: if $\labfunc(v)=\labfunc(w)$ 
then $\partial(v)=\partial(w)$.
 The translation $\trealasr$ 
is defined up to such labelling functions: 
$\trealasr(\pnone,\labfunc)$ is the graph $\sgone \in \grasr$ obtained in the following way:}
\begin{varitemize}
\item 
  Replace nodes $R_{\multimap}$ (resp.  $L_{\multimap}$) by nodes $\lambda$ (resp. $@$),
\item 
  Remove boxes and nodes $L_!$, $R_!$, $L_{\forall}$, $R_{\forall}$, $L_{\mu}$, $R_{\mu}$,
\item 
  Replace each contraction node $v$ with a fan-in with index $\labfunc(v)$. 
\end{varitemize}
We denote by  $|\labfunc|$  the cardinality of the image of the labelling function $\labfunc$.

\condinc{We say a labelling function $\labfunc$ for the proof-net $\pnone$ 
is \textit{compatible with depths} iff $\labfunc(v)=\labfunc(w)$ 
implies $\partial(v)=\partial(w)$. From now on
we will consider only labelling functions which are compatible with depths.}
{}
Note that in a proof-net reduction step $\pnone \rweal \pntwo$, each node of
$\pntwo$ comes from a unique node of $\pnone$; a labelling function $\labfunc$ for
$\pnone$ then induces in a natural way a labelling function
for $\pntwo$, that we will also write
$\labfunc$. By the stratification property of \EAL, if
$\labfunc$ is compatible with depths for $\pnone$, then so it is for
$\pntwo$.

In previous works on light logics and optimal reduction, two particular translations 
of proof-nets have been considered:
\begin{varitemize}
\item 
  The \emph{level translation}, LT: the labelling function $\labfunc$ is the one defined 
  by the depths, that is $\labfunc(v)=\partial(v)$.
\item 
  The \emph{distinct labelling translation}, DLT:
\condinc{ the labelling function $\labfunc$
  is the discrete one (each contraction node has a different index),}
{$\labfunc$ assigns a distinct index to each contraction node.}
\end{varitemize}
\condinc{
Observe that the second translation offers the advantage of simplicity,
since it does not need the information provided by boxes in $\pnone$. The
first translation, on the other hand, has the advantage that it
minimizes the number of indices used to annotate fans in the abstract
sharing graph. We will show that these two translations, as well as any one 
based on a labelling compatible with depths, is sound and complete for 
beta-reduction. For this purpose we will use as tool a specific 
\textit{context semantics}.
}
{
Note that the second translation  does not need the information provided by boxes in $\pnone$. The
first translation, on the other hand, has the advantage that it
minimizes the number of indices used to annotate fans.
The properties that we will prove in the rest of the paper
are valid for all these translations.
}
We give on Fig.\ref{fig:examplesg}(a) an abstract sharing graph 
  that will serve as running example. It is obtained as the DLT of a proof-net
corresponding to a derivation of $f:!(A \multimap A) \multimap !(A \multimap A) \multimap B, g: !(A \multimap A) \vdash u: B$,
where $u=(\lambda x. f \; x \; x)(\lambda z. g \; z)$. In  Fig. \ref{fig:examplesg}(b)
we give its normal form for $\rwasr$.
\begin{figure}
\begin{center}
  \subfigure[]
  {
  \begin{minipage}[c]{116pt}
    \centering\scalebox{.9}{\epsfbox{figure.85}}
  \end{minipage}
  }
  \hspace{10pt}
  \subfigure[]
  {
  \begin{minipage}[c]{87pt}
    \centering\scalebox{.9}{\epsfbox{figure.84}}
  \end{minipage}
  }
\caption{Example.}
\label{fig:examplesg}
\end{center}
\end{figure}

The concepts of principal port, direct path, simple path, etc. can be
easily transferred from proof-nets to sharing graphs. The number of
maximal paths in a cut-free sharing graph is bounded:
\begin{lemma}
Let $\sgone$ be a cut-free sharing graph and let $e$ be one
of its free ports. Then there are at most $|\sgone|+1$ maximal
direct paths in the form $e=e_1,\ldots,e_n$.
\end{lemma}
\condinc{
\begin{proof}
Consider any such maximal direct path $e=e_1,\ldots,e_n$
and the corresponding sequence of nodes $v_1,\ldots,v_{n-1}$.
Since there are no cuts in $\sgone$, there cannot
be any $e_i$ which is principal for both $v_{i-1}$ and $v_i$.
This implies $e_1,\ldots,e_n$ has a very constrained
structure: there is $1\leq j\leq n$ such that:
\begin{varitemize}
\item
  $e_i$ is principal for $v_i$ whenever $1\leq i<j$.
\item
  $e_j$ is not principal for $v_{j-1}$ (if $j>1$) nor for
  $v_j$ (if $j<n$).
\item
  $e_i$ is principal for $v_{i-1}$ whenever $j< i\leq n$
\end{varitemize}  
So, each such path can be divided into three parts. Now,
the third part of the path, namely $e_{j+1},\ldots,e_n$,
is completely determined by the first two parts, namely
$e_1,\ldots,e_j$. But since $e_1$ is always equal to $e$,
there are at most $|G|+1$ paths in this form, because 
every binary tree with $n$ nodes has at most $n+1$ 
leaves. This concludes the proof.
\end{proof}}{}
Now we want to bound the complexity of this rewriting procedure
and show that it is sound and complete.
\section{Context Semantics}

\condinc{\subsection{Interpretation of Proof-Nets and Sharing Graphs}
}{}
 \textit{Context semantics} will be the tool for showing
soundness of sharing graph reduction (following \cite{GAL92}).
A \textit{context} can be seen as a token carrying a piece of information
and travelling around the net \cite{DanosRegnier99}.
As we are considering a more constrained setting than 
 \cite{GAL92,girard89lc,danos95advances} the contexts can
be presented as tuples, as in \cite{BaillotPedicini01}. This reflects
the stratified structure of \EAL\  proof-nets.

\condinc{
\begin{definition}[Elementary contexts]\label{def:elementary-contexts}
An \emph{elementary context} $C$ of length $k$ is a tuple of stacks
$S_1,\ldots,S_k,T$ over the alphabet $\{\PP,\QQ\}$.  Stacks $S_i$ are
called \emph{exponential stacks}, stack $T$ is called
\emph{multiplicative stack}.  $\varepsilon$ denotes the empty stack, $xS$
denotes the stack obtained by pushing $x$ on $S$ and $ST$ denotes the
concatenation of stacks $S$ and $T$. The partial order $\sqsubseteq$
on stacks is the prefix order. We also denote by  $\sqsubseteq$ the pointwise order on the product of stacks.
Finally $\sqsubseteq_m$ will denote the order on elementary contexts defined by identity
on the exponential stacks $S_i$ ($1\leq i \leq k$) and  $\sqsubseteq$ on the multiplicative
stack $T$.
\end{definition}
}{
\begin{definition}[Elementary contexts]\label{def:elementary-contexts}
An \emph{elementary context} $C$ of length $k$ is a tuple of stacks
$S_1,\ldots,S_k,T$ over the alphabet $\{\PP,\QQ\}$.  Stacks $S_i$ are
called \emph{exponential stacks}, stack $T$ is called
\emph{multiplicative stack}.
\end{definition}
We denote by $\varepsilon$ the empty stack,
 $xS$ the result of  pushing $x$ on $S$ and by $SS'$ the
concatenation of $S$ and $S'$. Let $\sqsubseteq$
be the prefix ordering on stacks. We also denote by  $\sqsubseteq$ the pointwise order on the product of stacks.
Finally $\sqsubseteq_m$ will denote the order on elementary contexts defined by identity
on the exponential stacks $S_i$ and  $\sqsubseteq$ on the multiplicative
stack $T$.
}

\begin{definition}[Valid paths]\label{def:valid-paths}
  Let $\pnone$ be in  $\greal$   and $\labfunc$ a
  labelling function, with $k=|\labfunc|$.  
  \begin{varitemize}
    \item
      A \emph{context} of $\pnone$ relative to $\labfunc$ is a pair $(p,C)$
      where $p$ is an edge of $\pnone$ and $C$ is an elementary context 
      of length $k+1$. 
    \item
      The binary relation $\sim$ on contexts is defined by symmetrically
      closing the rules in Table~\ref{table:context-semantics-pn} and adding for
      the other (binary) nodes the rule acting as identity on the elementary context 
      (no rule for the $W$ node).
    \item
      A direct path $e_1,\ldots,e_n$ in $\pnone$ is \emph{valid} with respect 
      to two elementary contexts $C_1$ and $C_n$ iff the nodes along the path transform the 
      context $(e_1,C_1)$ into $(e_n,C_n)$. More precisely, there must be
      elementary contexts $C_2,\ldots,C_{n-1}$ such that
      $\cs{(e_i,C_i)}{(e_{i+1},C_{i+1})}$ whenever $1\leq i<n$.
      Then we write $\csor{(e_1,C_1)}{(e_n,C_n)}$ and say the path
      $e_1,\ldots,e_n$ is \emph{persistent}.
    \end{varitemize}
\end{definition}
\begin{table*}
\begin{center}
\caption{Context Semantics for Proof-nets}\label{table:context-semantics-pn}
\vspace{10pt}
\begin{tabular}{|c|c|}\hline\hline
\begin{minipage}[c]{2.5cm}
\vspace{.3cm}
\centering\scalebox{0.60}{\epsfbox{figure.26}}\\
\vspace{.3cm}
\end{minipage}
 &
\begin{minipage}[c]{12cm}
\begin{eqnarray*}
\csea{(e,(S_1,\ldots,S_{|\labfunc|},\PP T))}{(f,(S_1,\ldots,S_{|\labfunc|},T))}\\
\csea{(e,(S_1,\ldots,S_{|\labfunc|},\QQ T))}{(g,(S_1,\ldots,S_{|\labfunc|},T))}\\
\end{eqnarray*}
\end{minipage}
\\ \hline
\begin{minipage}[c]{2.5cm}
\vspace{.3cm}
\centering\scalebox{0.60}{\epsfbox{figure.27}}\\
\vspace{.3cm}
\end{minipage}
 &
\begin{minipage}[c]{12cm}
\begin{eqnarray*}
\csea{(e,(S_1,\ldots,S_{|\labfunc|},\PP T))}{(f,(S_1,\ldots,S_{|\labfunc|},T))}\\
\csea{(e,(S_1,\ldots,S_{|\labfunc|},\QQ T))}{(g,(S_1,\ldots,S_{|\labfunc|},T))}\\
\end{eqnarray*}
\end{minipage}
\\ \hline
\begin{minipage}[c]{2.5cm}
\vspace{.3cm}
\centering\scalebox{0.60}{\epsfbox{figure.30}}\\
\vspace{.3cm}
\end{minipage}
 &
\begin{minipage}[c]{12cm}
\begin{eqnarray*}
\csea{(e,(S_1,\ldots,S_{\labfunc(v)-1},\PP S_{\labfunc(v)},S_{\labfunc(v)+1},
\ldots,S_{|\labfunc|},T))}{(f,(S_1,\ldots,S_{|\labfunc|},T))}\\
\csea{(e,(S_1,\ldots,S_{\labfunc(v)-1},\QQ S_{\labfunc(v)},S_{\labfunc(v)+1},
\ldots,S_{|\labfunc|},T))}{(g,(S_1,\ldots,S_{|\labfunc|},T))}\\
\end{eqnarray*}
\end{minipage}
\\ \hline\hline
\end{tabular}
\end{center}
\vspace{-5mm}

\end{table*}
\begin{table*}
\begin{center}
\caption{Context Semantics for Sharing Graphs}\label{table:context-semantics-sg}
\vspace{10pt}
\begin{tabular}{|c|c|}\hline\hline
\begin{minipage}[c]{2.5cm}
\vspace{.3cm}
\centering\scalebox{0.90}{\epsfbox{figure.75}}\\
\vspace{.3cm}
\end{minipage}
 &
\begin{minipage}[c]{11cm}
\begin{eqnarray*}
\csea{(e,(S_1,\ldots,S_n,\PP T))}{(f,(S_1,\ldots,S_n,T))}\\
\csea{(e,(S_1,\ldots,S_n,\QQ T))}{(g,(S_1,\ldots,S_n,T))}\\
\end{eqnarray*}
\end{minipage}
\\ \hline
\begin{minipage}[c]{2.5cm}
\vspace{.3cm}
\centering\scalebox{0.90}{\epsfbox{figure.76}}\\
\vspace{.3cm}
\end{minipage}
 &
\begin{minipage}[c]{11cm}
\begin{eqnarray*}
\csea{(e,(S_1,\ldots,S_n,\PP T))}{(f,(S_1,\ldots,S_n,T))}\\
\csea{(e,(S_1,\ldots,S_n,\QQ T))}{(g,(S_1,\ldots,S_n,T))}\\
\end{eqnarray*}
\end{minipage}
\\ \hline
\begin{minipage}[c]{2.5cm}
\vspace{.3cm}
\centering\scalebox{0.90}{\epsfbox{figure.77}}\\
\vspace{.3cm}
\end{minipage}
 &
\begin{minipage}[c]{11cm}
\begin{eqnarray*}
\csea{(e,(S_1,\ldots,S_{i-1},\PP S_i,S_{i+1},\ldots,S_n,T))}{(f,(S_1,\ldots,S_n,T))}\\
\csea{(e,(S_1,\ldots,S_{i-1},\QQ S_i,S_{i+1},\ldots,S_n,T))}{(g,(S_1,\ldots,S_n,T))}\\
\end{eqnarray*}
\end{minipage}
\\ \hline\hline
\end{tabular}
\vspace{-5mm}
\end{center}
\end{table*}
\begin{definition}[Context semantics]\label{def:context-semantics}
Given a proof-net $\pnone$ of \EAL\  and a labelling function
$\labfunc$, the context semantics
$\conseman{\pnone}{\labfunc}$ of $\pnone$ is the set of pairs
$((e,C),(f,D))$ such that $e$ and $f$ are conclusion
edges of $\pnone$ and $\csor{(e,C)}{(f,D)}$.
\end{definition}
\condinc{
To simplify the notation we will sometimes omit the $\labfunc$ and 
write $\consem{\pnone}$ instead of $\conseman{\pnone}{\labfunc}$.
}
{
We will sometimes write simply $\consem{\pnone}$ instead of $\conseman{\pnone}{\labfunc}$.
}
\condinc{
Notice that as the transitions are deterministic (see
Table~\ref{table:context-semantics-pn}) and as when reaching a conclusion
no transition is possible anymore, if $\csor{(e,C)}{(f,D)}$ and 
$\csor{(e,C)}{(g,E)}$ are both in $\conseman{\pnone}{\labfunc}$ 
then $f=g$ and $D=E$. Therefore the
context semantics of $\pnone$ can be seen as a (partial) function on
contexts. Notice, however, that there can be two essentially different
reasons why the context semantics is undefined on $(e,C)$:
\begin{varitemize}
\item
  There could be finitely many valid paths starting in $e$ which are all valid with respect
  to $C$ and some context $D$, but none of them ends in a conclusion.
\item
  There are arbitrary long valid paths starting in $e$ which are all valid with respect
  to $C$ and some context $D$; this means there is not any context $(g,E)$ such
  that $g$ is a conclusion edge and $\csor{(e,C)}{(g,E)}$.
\end{varitemize}
}
{
Notice that as the transitions are deterministic  and as when reaching a conclusion
no transition is possible anymore, if $\csor{(e,C)}{(f,D)}$ and 
$\csor{(e,C)}{(g,E)}$ are both in $\conseman{\pnone}{\labfunc}$ 
then $f=g$ and $D=E$. Therefore the
context semantics of $\pnone$ induces a (partial) function on
contexts. Note, however, that there can be two essentially different
reasons why the context semantics is undefined on $(e,C)$:
\begin{varitemize}
\item
The current context could get ``stuck'' at some node: there is no possible
transition, 
\item
 The current context could travel indefinitely without reaching a conclusion.
\end{varitemize}
}
However we will see in Section~\ref{sect:acyclicity} that this second possibility
is guaranteed never to occur for proof-nets.

Given a sharing graph $\sgone$ and a partition $\labfunc$ of its fan
nodes, its contexts and context semantics $\consem{\sgone}$ are
defined similarly to that of proof-nets (Table
\ref{table:context-semantics-sg}). It is then clear that the context
semantics is preserved by the  translation from proof-nets to
sharing graphs:
 
\begin{proposition}\label{lem:c.s.-preservation-wrt-sg-translation}
Let $\pnone$ be an \EAL\  proof-net and $\labfunc$ a partition of its contraction nodes, then 
$\conseman{\pnone}{\labfunc}=\consem{\trealasr(\pnone,\labfunc)}$.
\end{proposition}


We give some examples of contexts in the context semantics of the sharing graph
from Fig.\ref{fig:examplesg}(a):
$$
\begin{array}{rclrcl}
 \csora{(f, \varepsilon,\PP\QQ)}{(g,\PP,\QQ)}; & \csora{(f, \varepsilon,\QQ\PP \QQ)}{(g,\QQ,\QQ)};\\
 \csora{(e, \varepsilon,\varepsilon)}{(f,\varepsilon,\QQ \QQ)}; & \csora{(g, \PP ,\PP)}{(f,\varepsilon,\PP\PP)}; \\
 \csora{(g, \QQ ,\PP)}{(f,\varepsilon,\QQ\PP\PP)}. &
\end{array}
$$
The path corresponding to the first of these contexts is represented on
 Fig.\ref{fig:examplesg}(a) by a dashed line.

If $P$ is a set of contexts, $P^-$ denotes the subset of $P$ including
only minimal elements (with respect to $\sqsubseteq$).
When traversing any node in a sharing graph $\sgone$, only one
particular stack of the underlying context can be modified. Two
nodes $u$ and $v$ have the same sort (formally, $\type{u}=\type{v}$)
iff they can modify the same stack. For instance $@$ and
$\lambda$ nodes have the same sort.
Given  a node $u$, $\epsi{u}$ is the set of contexts
whose stack corresponding to $u$ is $\varepsilon$.

\begin{lemma}[Monotonicity]
Suppose $e_1,\ldots,e_n$ is a direct path valid with 
respect to $C_1,\ldots,C_k,T$ and
$D_1,\ldots,D_k,S$. Moreover, suppose that
$E_1,\ldots,E_k,U$ are stacks. Then
$e_1,\ldots,e_n$ is valid with respect
to $C_1E_1,\ldots,C_kE_k,TU$
and $D_1E_1,\ldots,D_kE_k,SU$.
\end{lemma}
\condinc{
\begin{proof}
By induction on $n$.
\end{proof}
\begin{proposition}[Minimality]\label{prop:minimality}
For every persistent path $e_1,\ldots,e_n$ there are elementary contexts
$C$ and $D$ such that whenever $e_1,\ldots,e_n$ is valid with respect
to $E$ and $F$, $C\sqsubseteq E$ and $D\sqsubseteq F$.
\end{proposition}
\begin{proof}
By induction on $n$.
\end{proof}}{}
\subsection{Reduction and Context Semantics}
Now we consider the behaviour of the context semantics with respect to
the reduction of proof-nets and sharing graphs. Let us start with the 
latter case, which is easier.

\condinc{
Take a look at the rewriting rules for sharing graph. If we focus on
the edges involved, we can observe that: 
\begin{varitemize}
\item
  The annihilation rewriting steps erase one edge, namely the
  cut. 
\condinc{The other four edges involved have \emph{residuals} which are defined in the
  usual way. The edges which are not directly involved in the
  rewriting have trivially defined residuals.
}
{
The other four edges involved, as well as those not directly involved
in the rewriting, have \emph{residuals} which are defined in the
  usual way.
}
 No edge is created. 
\item
  The copying rewriting steps erases one edge but creates another
  four edges, which are called the \emph{edges created in the
  rewriting step}. The cut in the redex has no residual.
\end{varitemize}
Let $\sgone$ be a sharing graph and let $E$ be a subset of the edges of $\sgone$. 
The direct path $e_1,\ldots,e_n$ in $\sgone$ is said to be 
\emph{long enough for $E$} iff $e_1,e_n\notin E$.
\begin{lemma}[Preservation of Long-Enough Paths]\label{lemma:preservation}
  Suppose $\sgone$ is a sharing graph, $\sgone\rwasr\sgtwo$ by
  firing a cut $e$. Then:
  \begin{varitemize}
  \item
    If a direct path $e_1,\ldots,e_n$ in $\sgone$
    is long enough for $\{e\}$ and valid for $C$ and $D$, then
    there is a direct path $g_1,\ldots,g_m$ in $\sgtwo$ valid
    for $C$ and $D$ such that $g_1$ is the residual of $e_1$ 
    and $g_m$ is the residual of $e_n$.
  \item
    If a direct path $g_1,\ldots,g_m$ in $\sgtwo$
    is long enough for the set of edges created in the 
    rewriting step and valid for $C$ and $D$, then there 
    is a direct path $e_1,\ldots,e_n$ in $\sgone$ valid
    for $C$ and $D$ such that $g_1$ is the residual of $e_1$ 
    and $g_m$ is the residual of $e_n$
  \end{varitemize}
\end{lemma}
\begin{proof}
Consider the rules of Figure \ref{figure:asrrw} and observe that in each case of
rewriting step the context semantics partial function of the subgraph
concerned is unchanged.
\end{proof}
}{Since sharing graph reduction is local, all valid paths starting and ending
with a free port are preserved during reduction. Consequently:
}
\begin{proposition}\label{lem:c.s.-preservation-wrt-sg-reduction}
Let $\sgone$ be a sharing graph and $\sgone\rwasr\sgtwo$ then 
$\consem{\sgone}=\consem{\sgtwo}$.
\end{proposition}

\condinc{
\begin{proof}
Just observe that any conclusion-to-conclusion valid path in $\sgone$
is long enough for any cut, while any conclusion-to-conclusion valid path
in $\sgtwo$ is long enough for the set of edges created in the rewriting
step. The thesis follows easily from Lemma~\ref{lemma:preservation}.
\end{proof}}{}

As to proof-nets the situation is more delicate. It
is well-known that geometry of interaction or context semantics are
usually not preserved by general proof-net reduction~\cite{girard89lc,danos95advances,mairson02fsttcs}. 
To deal with this problem  we define a
partial order $\preccurlyeq$ on context functions.
Context semantics will be preserved up to $\succcurlyeq$ but that
 will be sufficient to obtain a soundness result with respect to
lambda-calculus. 

\condinc{
\begin{definition}\label{def:order-context-semantics}
 Let $f$, $g$ be two partial functions on contexts. Then $f \preccurlyeq g$ iff 
 for any context $p,C$  we have:
 \begin{varenumerate}
 \item 
   If $f(p,C)$ is defined, then so is $g(p,C)$, and $g(p,C)=f(p,C)$,
 \item\label{caseb}
   If $f(p,C)$ is undefined then either:
   \begin{varvarenumerate}
   \item 
     $g(p,C)$ is undefined,
   \item\label{subcaseb} 
     or $f(p,D)$ is undefined whenever $D\sqsupseteq_m C$.
   \end{varvarenumerate}
 \end{varenumerate}
\end{definition}
 The point in subcase \ref{caseb}.ii is that $f(p,C)$ is undefined,
but it is not merely because of a lack of information in the multiplicative
stack, since no increase of information on this stack can trigger an
answer. The behaviour of $f$ on such input is in fact irrelevant for the
read-back process that we will define, so the definition of $\preccurlyeq$ does not
require anything on $g(p,C)$ in this case.
}
{
\begin{definition}\label{def:order-context-semantics}
 Let $f$, $g$ be two partial functions on contexts. Then $f \preccurlyeq g$ iff 
 for any context $p,C$  we have:
 \begin{varenumerate}
 \item 
   If $f(p,C)$ is defined, then $g(p,C)=f(p,C)$,
 \item\label{caseb}
   If $f(p,C)$ is undefined then either:
   \begin{varvarenumerate}
   \item 
     $g(p,C)$ is undefined,
   \item\label{subcaseb} 
     or $f(p,D)$ is undefined whenever $D\sqsupseteq_m C$.
  \end{varvarenumerate}
 \end{varenumerate}
\end{definition}
 The point in \ref{caseb}.ii is that $f(p,C)$ is undefined,
but it is not merely because of a lack of information in the multiplicative
stack, since no increase of information on this stack can trigger an
answer. The behaviour of $f$ on such input is in fact irrelevant for the
read-back process that we will define. 
}

\begin{lemma}
 The relation $\preccurlyeq $ is a partial order.
\end{lemma}
\condinc{
\begin{proof}
 The non-obvious fact is whether this relation is transitive. Assume we have $f \preccurlyeq g$
 and  $g \preccurlyeq h$. Given a context $(p,C)$ and a stack $T_0$ we will denote
 by $(p,C)::T_0$ the context obtained from $p,C$ by replacing the multiplicative stack $T$ of $C$ by
 $T T_0$.
 
 Take a context $(p,C)$: if $f(p,C)$ is defined, then so are $g(p,C)$
 and $h(p,C)$, and we have $h(p,C)=f(p,C)$. Otherwise if $f(p,C)$ is
 undefined we have 2 subcases to consider. First, if for any stack
 $T_0$, $f((p,C)::T_0)$ is undefined, then the condition is
 fulfilled. Otherwise there exists a $T_0$ such that $f((p,C)::T_0)$
 is defined, and $g(p,C)$ is undefined; then $g((p,C)::T_0)$ is
 defined. As $g \preccurlyeq h$ we deduce that $h(p,C)$ is undefined
 and the condition is fulfilled. Therefore $f \preccurlyeq h$.
\end{proof}}{}
 \condinc{Now we can state the property of context semantics w.r.t. proof-net
 reduction:}
{Now we have:}
\begin{proposition}\label{lem:c.s.-weak-preservation-wrt-pn-reduction}
Let $\pnone$ be an \EAL\  proof-net and $\pnone\rweal \pntwo$ then
$\consem{\pnone} \succcurlyeq \consem{\pntwo}$.
\end{proposition}

\condinc{
\begin{proof}
Consider one step of reduction $\pnone\rweal \pntwo$. We want to define
a map $\phi$ sending each edge $e$  of $\pntwo$ to an edge of $\pnone$ of
same type. First,
every conclusion $e$ of $\pntwo$ is naturally associated to a conclusion
$e'$ of $\pnone$, and we define $\phi(e)=e'$.  For the other edges
we have to distinguish among the different cases of reduction steps
 of Figure \ref{figure:ealpnrw};
 we only describe the map on the edges involved in the reduction
step, for the other edges it is defined as expected. Let us consider
the various steps (using the notations of  Figure \ref{figure:ealpnrw}): 
\begin{itemize}
\item $\multimap$ reduction step: the edge of $\pntwo$ of type $A$ (resp. $B$) is mapped to
the $A$ edge of $\pnone$ incident to the $R_{\multimap}$ node (resp. the $B$ edge of $\pnone$
 incident to the $L_{\multimap}$ node).
\item Box-box reduction step: the $B$ edge inside the box of $\pntwo$ is mapped
to the $B$ edge of $\pnone$ incident to the $R_{!}$ node; the other edges are mapped
in the natural way.
\item Contraction step: for each $X$ node in $\pntwo$ created in the reduction step,
the three incident edges with type $!A_i$ are mapped to the $!A_i$
edge of $\pnone$ incident to $L_{!}$; each edge in the two boxes of $\pntwo$ is mapped
to the corresponding edge in the box of $\pnone$; the $!B$ edge of the left (resp. right) box
is mapped to the left (resp. right) non-principal edge of $X$ in $\pnone$.
\item $\mu$ reduction step: the $A[\mu \alpha. A \slash \alpha]$ edge of $\pntwo$
is mapped to the  $A[\mu \alpha. A \slash \alpha]$ edge of $\pnone$ incident to $R_{\mu}$.
\item $\forall$ reduction step: as in the $\mu$ reduction step.
\end{itemize}


 We now define a map from contexts of $\pntwo$ to contexts of $\pnone$, sending
a context $(p',C')$ to a context $(p,C)$ of $\pnone$, with $p=\phi(p')$,
and that we also denote as $\phi$. If the reduction step considered is
any step but the contraction step, then $\phi$ is simply the
identity. In the contraction case: denote by $v$ the contraction node
in $\pnone$ involved in this step, by $b$ the box in $\pnone$ to be duplicated,
and by $b_1$, $b_2$ its two copies in $\pntwo$. Let $i=\labfunc(v)$. Take a context $(p',C')$ in
$\pntwo$: if $p'$ is not in one the $b_j$ boxes nor one of their premises,
 then $\phi(p',C')= (\phi(p'),C')$. If $p'$ is in $b_1$
(resp. $b_2$), or one of its $!A_i$ premises, then
$\phi(p',C')= (\phi(p'),C)$, where $C$ is obtained from $C'$ by
replacing the $i$th stack $S_i$ by $\sf{p}S_i$ (resp. $\sf{q}S_i$).

Let us denote by $\sim^{\ast}$ the transitive and reflexive closure
of the $\sim$ relation in $\pnone$.
\begin{lemma}\label{lem:transitionlifting}
Let $\pnone \rweal \pntwo$. If $(e,C) \sim(f,D)$ is a transition of $\pntwo$,
then $\phi(e,C)\sim^{\ast}\phi(f,D)$ is obtained by a (possibly empty) sequence of transitions
in $\pnone$. 
\end{lemma}
\begin{proof}
 One can check it by examining for each case of reduction step in
 Figure \ref{figure:ealpnrw} the various possible transitions in
$\pntwo$. Let us just examine here one case for the example.

Consider a contraction reduction step, denote $v$ the contraction
node involved in $\pnone$, and take a transition $(e,C) \sim(f,D)$
in $\pntwo$ corresponding to a node $v'$ inside one of the two
boxes, say the left one $b_1$. Assume for
instance $v'$ is a contraction node (the other cases are easier).
 Denote $i=\labfunc(v)$ and $j=\labfunc(v')$. We have 
$\partial(v')\geq \partial(v)+1$, therefore 
as $\labfunc$ is
compatible with depths we get  $i\neq j$.
 Then by definition of $\phi$: 
$\phi(e)$ and $\phi(f)$ are incident to a contraction node $v''$ in $\pnone$.
Moreover $\labfunc(v'')=j$. Therefore 
$\phi(e,C)$ (resp. $\phi(f,D)$ ) has same $j$-th stack as $(e,C)$
(resp. $(f,D)$ ) (only the $i$-th stack has been modified) and it 
follows that $\phi(e,C) \sim \phi(f,D)$ is a transition of $\pnone$.
\end{proof}

 Consider a valid path in $\pntwo$ and a corresponding sequence of contexts
 $s=(p_1,C_1), \dots, (p_n,C_n)$ 
following the transitions of this proof-net. By using Lemma \ref{lem:transitionlifting}
for each of its transitions and concatenating together the paths obtained
in $\pnone$, one obtains a path in $\pnone$ which is direct, and valid because it
transforms context $\phi(p_1,C_1)$ into context  $\phi(p_n,C_n)$.

 It follows that if the function
$\consem{\pntwo}$ is defined on a context $(p_0,C_0)$, then so is
$\consem{\pnone}$ and we have $\consem{\pntwo} (p_0,C_0)= \consem{\pnone}
(p_0,C_0)$.

 Now, assume $\consem{\pntwo} (p_0,C_0)$ is undefined. Let $(p_0,C_0),
\dots, (p_n,C_n)$ be the corresponding sequence of contexts in $\pntwo$,
with $(p_n,C_n)$ not admitting any further transition and such that $p_n$ is not a conclusion.
 As just said
there is a valid path in $\pnone$ with  sequence of contexts  containing
 (as subsequence) $\phi(p_0,C_0), \dots,
\phi(p_n,C_n)$.  If we are in the case of a non-contraction reduction
step, then as $\phi$ acts as the identity on elementary contexts 
we have $\phi(p_n,C_n)=(\phi(p_n),C_n)$. As in $\pntwo$ the context $(p_n,C_n)$ does not admit 
any further transition, it is the same for  $\phi(p_n,C_n)$ in $\pnone$. Moreover 
$\phi(p_n)$ is not a conclusion,  hence $\consem{\pnone} (p_0,C_0)$ is
undefined. Therefore in this case we have $\consem{\pntwo} \preccurlyeq \consem{\pnone}$.

 In the case where the reduction step is a contraction one we keep the
same notations for the involved boxes and contraction node defined
before. Let us consider again the sequence in $\pntwo$ $(p_1,C_1), \dots,
(p_n,C_n)$.  As in $\pntwo$ the transition on $(p_n,C_n)$ is not defined,
this context is entering the principal port of a node $v$ (see Table
\ref{table:context-semantics-pn}). We have two cases:
\begin{varitemize}
\item If $v$ is a $R_{\multimap}$ or $L_{\multimap}$ node, then this means that the
multiplicative stack $T$ of $C_n$ is $\varepsilon$, so by definition the
multiplicative stack of $\phi(p_n,C_n)$ is also empty and thus in $\pnone$
no transition is possible for $\phi(p_n,C_n)$. Therefore in this case
$\consem{\pnone} (p_0,C_0)$ is undefined.
\item If $v$ is an $X$ node (contraction), let $k=\labfunc(v)$. Then as the transition is
undefined, the $k$th stack $S_k$ of $C_n$ is $\varepsilon$. Consider
 $D$ such that $C_1 \sqsubseteq_m D$. Let $T$ be the multiplicative stack of $C_1$.
Then there exists $T_0$ such that $D$'s multiplicative stack is $TT_0$.
Let us now denote by $(p,C::T_0)$ the context obtained from
$(p,C)$ by replacing the multiplicative stack $T$ of $C$ by
$TT_0$.  Then $(p_1,C_1::T_0)= (p_1,D)$. The following sequence is obtained by consecutive
transitions in $\pntwo$: $(p_1,C_1::T_0), \dots, (p_n,C_n::T_0)$. Moreover
 $(p_n,C_n::T_0)$ has an empty $k$th stack; hence just as $(p_n,C_n)$, the context 
$(p_n,C_n::T_0)$ has no possible transition in $\pnone$. It follows that $\consem{\pntwo}
(p_0,C_0::T_0)$ is undefined.  Therefore we are in the case
\ref{subcaseb}(ii) of Definition \ref{def:order-context-semantics}. 
\end{varitemize}
 So we can conclude that $\consem{\pntwo}\preccurlyeq \consem{\pnone}$.
\end{proof}}{}
\subsection{Acyclicity}\label{sect:acyclicity}
We now describe properties of valid paths in the proof-nets and
sharing graphs we are dealing with.
\begin{proposition}[Finiteness of valid paths for proof-nets]\label{pnpathsfiniteness}
Let $\pnone$ be an \EAL\  proof-net. Then there exists
an integer $k$ such that for any valid path $e_1, \dots, e_n$ we have 
$n\leq k$.
\end{proposition}
\begin{proof}
 This is proved in \cite{BaillotPedicini01} for \EALso, and the proof can be
easily adapted to \EAL\, using the fact that \EAL\ is strongly
normalising.
\end{proof}

\condinc{
A \emph{cycle} is a direct path $e_1,\ldots,e_n$ such
that:
\begin{varitemize}
\item
  $e_1=e_n$;
\item
  $n\geq 2$;
\item
  $e_1,\ldots,e_n$ is valid with respect to $C=C_0,\ldots,C_k,T$ and
  $D=D_0,\ldots,D_k,S$;
\item
  For every $0\leq i\leq k$, either $C_i\sqsubseteq D_i$
  or $D_i\sqsubseteq C_i$;
\item
  Either $T\sqsubseteq S$ or $S\sqsubseteq T$.
\end{varitemize}
}
{
A \emph{cycle} is a direct path $e_1,\ldots,e_n$ with $n\geq 2$ such
that   $e_1=e_n$ and :
\begin{varitemize}
\item
  $e_1,\ldots,e_n$ is valid with respect to $C=C_0,\ldots,C_k,T$ and
  $D=D_0,\ldots,D_k,S$;
\item
  For every $0\leq i\leq k$, either $C_i\sqsubseteq D_i$
  or $D_i\sqsubseteq C_i$;
\item
  Either $T\sqsubseteq S$ or $S\sqsubseteq T$.
\end{varitemize}
}
\begin{proposition}[Acyclicity of Proof-Nets]\label{pnacyclicity}
If $\pnone$ is a proof-net, then its context semantics does not contain 
any cycle.
\end{proposition}
\begin{proof}
 Indeed if the proof-net
 $\pnone$ contained a cycle,
then by repeatedly composing it with itself one would get valid paths
of arbitrary length, which would contradict Proposition~\ref{pnpathsfiniteness}.
\end{proof}


\begin{proposition}\label{prop:sgpathsfiniteness}
Let $\pnone$ be an \EAL\  proof-net,
$\sgone=\trealasr(\pnone,\labfunc)$ and $\sgone\rwasr^*\sgtwo$ . Then
there exists an integer $k$ such that: for any valid path $e_1, \dots,
e_n$ of $\sgtwo$ we have $n\leq k$.
\end{proposition}
\condinc{
\begin{proof}
 First, the statement holds for the paths of $\sgone$ itself because
of Prop. \ref{pnpathsfiniteness} and of the fact that any valid
 path of $\sgone$ can be lifted back
to a valid path of $\pnone$  of same length or longer. Then consider
$\sgtwo$ obtained from $\sgone$ by one step of $\rwasr$. Using Lemma \ref{lemma:preservation}
one can show that if $\sgtwo$ has valid paths of arbitrary length, then so has
$\sgone$, which yields a contradiction. Hence the property is valid for any $\sgtwo$ such that
 $\sgone\rwasr\sgtwo$.
\end{proof}
}{
\begin{proof}[sketch]
 The statement holds for the paths of $\sgone$ itself because
of Prop. \ref{pnpathsfiniteness} and of the fact that any valid
 path of $\sgone$ can be lifted back
to a valid path of $\pnone$  of same length or longer. Then for $\sgtwo$ obtained from $\sgone$
 by one step of $\rwasr$, one can show that  if $\sgtwo$ has valid paths of arbitrary length, then so has
$\sgone$, which yields a contradiction.
\end{proof}
}
\section{Complexity}\label{sect:complexity}
We study the complexity of sharing graph reduction by defining
a weight $W_G$ for any sharing graph $G$. The underlying idea
is the following: the weight of $G$ is the sum of the individual
weight of each $u\in V_G$, the latter being the number of possible
copies of $u$ that are produced during normalisation. We will 
relate the weight to the number of reduction steps of the sharing graph,
and then, for sharing graphs coming from \EAL (and \LAL), bound the weight
by using the properties of proof-nets.

Formally, the weight
of an edge $u$ will be defined as the number
of different valid paths $e_1,\ldots,e_n$ satisfying certain
additional constraints. First of all, $e_1$ must be $\pp{u}$.
Secondly, $e_n$ must be:
\begin{varitemize}
\item
  Either the principal edge of a node $v$ such that
  $\type{u}=\type{v}$.
\item
  Or an edge in $\fp{\sgone}\cup \wpo{\sgone}$.
\end{varitemize}
This way the weight of $u$ will be exactly the number of copies
of $u$ that will eventually appear during reduction of $G$.
 This can be characterized by context semantics \condinc{exploiting Proposition~\ref{prop:minimality}}{}:
\begin{definition}[Weight]
Let $\sgone\in\grasr$. Then:
\begin{varitemize}
  \item
    If $u$ is a node of $\sgone$, then $B_u$, $P_u$ and $E_u$ are sets of
    elementary contexts defined in Figure~\ref{fig:bupueu}.
    \begin{figure*}
      \begin{eqnarray*}
        B_u&=&\left\{C\;|\exists v,D.\big(\csor{(\pp{u},C)}{(\pp{v},D)}\big)
          \wedge (C\in\epsi{u})\wedge (D\in\epsi{v})
          \wedge(\type{u}=\type{v})\right\}\\
        P_u&=&\left\{C\;|\exists q,D.\big(\csor{(\pp{u},C)}{(q,D)}\big
          )\wedge(C\in\epsi{u})\wedge(q\in\fp{\sgone})\right\}\\
        E_u&=&\left\{C\;|\exists q,D.\big(\csor{(\pp{u},C)}{(q,D)}\big)
          \wedge(C\in\epsi{u})\wedge(q\in\wpo{\sgone})\right\}
      \end{eqnarray*}
      \caption{Definition of $B_u$, $P_u$ and $E_u$.}\label{fig:bupueu}
    \end{figure*}
  \item
    The weight $W_\sgone$ of $\sgone$ is defined as follows:
    $$
    W_\sgone=\sum_{u\in V_\sgone} (|B_u^-|+|P^-_u|+|E^-_u|-1).
    $$
    Notice that $W_\sgone$ can be either a natural number or
    $\omega$.
\end{varitemize}
\end{definition}
Notice how $1$ is subtracted from the sum $|B_u^-|+|P^-_u|+|E^-_u|$
when defining $W_\sgone$. This way, $W_\sgone$ always
decreases at any copying normalisation step, as we will see.
The weight of a cut-free sharing graph obtained by reducing 
another sharing graph coming from a proof-net is always null:
\begin{lemma}\label{lemma:sgcutfree}
If $\pnone\in\greal$ and $\trealasr(\pnone,\labfunc)\rwasr^*\sgone$
where $\sgone\in\grasr$ is a cut-free graph, then $W_{\sgone}=0$.
\end{lemma}
\begin{proof}
Consider any $u\in V_\sgone$ and any direct path starting from
$u$. This path is always simple, since we assume $\sgone$ to be
cut-free. Moreover, by Proposition~\ref{prop:sgpathsfiniteness}, 
we cannot go on forever building up the path. As a consequence, we will end up 
at an edge in $\fp{\sgone}\cup\wpo{\sgone}$. This, in particular,
implies that $|P^-_u|+|E^-_u|=1$, while $|B_u^-|=0$. The thesis follows
easily.
\end{proof}
\begin{lemma}\label{lemma:cutfreeasr}
If $\pnone\in\greal$ is a cut-free proof-net and $\labfunc$ is any
partition of its contraction nodes, then $W_{\trealasr(\pnone,\labfunc)}=0$.
\end{lemma}
\condinc{
\begin{proof}
Trivial, since $\trealasr(\pnone,\labfunc)$ is cut-free whenever $\pnone$ is
cut-free and, as a consequence, we can apply Lemma~\ref{lemma:sgcutfree}.
\end{proof}}{}
\condinc{}{Now, observe that annihilation rewriting steps leave 
$W_\sgone$ unchanged (while $|\sgone|$ decreases by $2$), whereas
copying rewriting steps make $W_\sgone$ decrease by $2$ (while $|\sgone|$ increases
by $2$). As a consequence:}
\begin{proposition}\label{prop:weightbound}
If $\pnone\in\greal$, $\sgone=\trealasr(\pnone,\labfunc)$,
$W_\sgone$ is finite and $\sgone\rwasr^n \sgtwo$, then $n\leq W_\sgone+|\sgone|/2$
and $|\sgtwo|\leq W_\sgone+|\sgone|$.
\end{proposition}
\condinc{
\begin{proof}
It is sufficient to observe that:
\begin{varitemize}
  \item
    Annihilation rewriting steps leave $W_\sgone$ unchanged, while $|\sgone|$ decreases
    by $2$.
  \item
    Copying rewriting steps make $W_\sgone$ decrease by $2$, while $|\sgone|$ increases
    by $2$.
\end{varitemize}
This implies, by Lemma~\ref{lemma:sgcutfree} that $W_G$ is finite and that
the total number of copying rewriting steps is $W_\sgone/2$. As a consequence, the
size of $\sgtwo$ is at most $W_\sgone+|\sgone|$. Moreover, the total number of 
annihilation rewriting steps is $(W_\sgone+|\sgone|)/2$. This completes the proof.
\end{proof}}
\condinc{}{Given a proof-net $\pnone\in\greal$ such that $\pnone\rweal\pntwo$, we can study the
difference $W_{\trealasr(\pnone,\labfunc)}-W_{\trealasr(\pntwo,\labfunc)}$. In particular, in the
case of the MLBL strategy, the difference can be tightly bounded, because the number
of paths that we ``lose'' at each reduction step can be properly bounded by an appropriate
function (with the same order of magnitude as the one from Theorem~\ref{th:pnbound}) applied
to $|\pnone|$. This implies the weight of the underlying sharing graph cannot decrease too much during 
MLBL proof-net reduction and, by Lemma~\ref{lemma:cutfreeasr} and Theorem~\ref{th:pnbound}, 
we get:}
\begin{proposition}\label{prop:complexeal}
For every natural number $n$, there is an elementary function 
$e_n:\N\rightarrow\N$ such that for every proof-net $\pnone\in\greal$,
$W_{\trealasr(\pnone)}\leq e_{\partial(\pnone)}(|\pnone|)$.
\end{proposition}
\condinc{
\begin{proof}
First of all, we know that for every natural number $n$, there are
elementary functions $f_n,g_n:\N\rightarrow\N$ such that for every
proof-net $\pnone\in\greal$ if $\pnone\rweal^m \pntwo$, then
$m\leq f_{\partial(\pnone)}(|\pnone|)$ and $|\pntwo|\leq g_{\partial(\pnone)}(|\pnone|)$.
We can build up $e_n$ by induction on $n$:
\begin{varitemize}
  \item
    $e_0(x)=0$ for every $x\in\N$. Indeed,
    let $\partial(\pnone)=0$. If $\pnone\rwlal \pntwo$ in the modified level-by-level strategy, 
    then $W_{\trealasr(\pnone)}=W_{\trealasr(\pntwo)}$ and, moreover, $W_{\trealasr(\pnone)}=0$ 
    whenever $\pnone$ is cut-free (by Lemma~\ref{lemma:cutfreeasr}).
  \item
    $e_n(x)=f_n(x)\cdot (e_{n-1}(g_n(x))+2g_n(x))$ for every $n\geq 1$. Indeed,
    let  $\partial(\pnone)=n\geq 1$. If $\pnone\rweal\pntwo$ in the MLBL strategy, then 
    $W_{\trealasr(\pnone)}-W_{\trealasr(\pntwo)}\leq e_{n-1}(|\pnone|)+2|\pnone|$. This because:
    \begin{varitemize}
    \item
      At any normalisation step other than copying, 
      $W_{\trealasr(\pnone)}=W_{\trealasr(\pntwo)}$, as we already pointed out.
    \item
      In the case of copying, we are in the situation depicted in Figure~\ref{figure:expmlbl}.
      \begin{figure}
        \begin{center}
          \begin{minipage}[c]{61pt}
            \centering\scalebox{0.5}{\epsfbox{figure.81}}
          \end{minipage}
          \begin{minipage}[c]{28pt}
            \centering $\rweal$
          \end{minipage}
          \begin{minipage}[c]{137pt}
            \centering\scalebox{0.5}{\epsfbox{figure.82}}
          \end{minipage}
        \end{center}
        \caption{}
        \label{figure:expmlbl}
      \end{figure}
      $W_{\trealasr(\pnone)}-W_{\trealasr(\pntwo)}$ can be bounded as follows:
      \begin{varitemize}
      \item
        Consider any $u\in V_\pnthree$ and any persistent path in $\pnone$
        starting from $u$. Any such path can be mimicked by at least one
        of the two copies $u_1$ and $u_2$ of $u$ appearing in
        $\pntwo$. In particular, if 
        the path stays inside $\pnthree$, than it can be mimicked by two
        paths starting in $u_1$ and $u_2$, while if the path exits
        from $\pnthree$, 
        it can be mimicked by exactly one path starting in either
        $u_1$ or $u_2$. By definition of $W_{\trealasr(\pntwo)}$, the
        contribution to the weight of these nodes decreases by at most
        $|\pnone|$.
      \item
        Consider the node $w\in V_\pnone$. Paths starting in $w$ cannot be mimicked
        by any in $V_\pntwo$. We claim, however, that there cannot be more than
        $p+1$ such paths, where $p$ is the size of the normal form
        of $\trlalasr(\pnthree)$. Indeed, all such paths can be seen as 
        maximal, persistent paths in $\pnthree$. By Proposition~\ref{prop:weightbound}.
        the size of the normal form of $\trlalasr(\pnthree)$ cannot be more
        than $e_{n-1}(|\pnone|)+|\pnone|$.
      \end{varitemize}
    \end{varitemize}
    As a consequence, since $W_{\trealasr(\pnone)}=0$ whenever $\pnone$ is cut-free 
    (again by Lemma~\ref{lemma:cutfreeasr}), we can iterate over
    the inequality $W_{\trealasr(\pnone)}-W_{\trealasr(\pntwo)}\leq e_{n-1}(|\pnone|)+2|\pnone|$
    obtaining $W_{\trealasr(\pnone)}\leq f_n(|\pnone|)(e_{n-1}(g_n(|\pnone|))+2g_n(|\pnone|))$ (since
    $g_n(|\pnone|)$ is a bound on the size of any reduct of $\pnone$).
  \end{varitemize}
  This concludes the proof.
\end{proof}}{}
\condinc{}{A similar result holds for \LAL:}
\begin{proposition}\label{prop:complexlal}
For every natural number $n$, there is a polynomial
$e_n:\N\rightarrow\N$ such that for every proof-net $\pnone\in\grlal$,
$W_{\trlalasr(\pnone)}\leq e_{\partial(\pnone)}(|\pnone|)$.
\end{proposition}
\condinc{
\begin{proof}
The proof is similar to the one for Proposition~\ref{prop:complexeal}.
First of all, we know that for every natural number $n$, there are
polynomials $f_n,g_n:\N\rightarrow\N$ such that for every
proof-net $\pnone\in\greal$ if $\pnone\rweal^m \pntwo$, then
$m\leq f_{\partial(\pnone)}(|\pnone|)$ and $|\pntwo|\leq g_{\partial(\pnone)}(|\pnone|)$.
We can build up $e_n$ by induction on $n$:
\begin{varitemize}
  \item
    $e_0(x)=0$ for every $x\in\N$. Indeed,
    let $\partial(\pnone)=0$. If $\pnone\rweal \pntwo$ in the modified level-by-level strategy, 
    then $W_{\trlalasr(\pnone)}=W_{\trlalasr(\pntwo)}$ and, moreover, $W_{\trlalasr(\pnone)}=0$ 
    whenever $\pnone$ is cut-free (by Lemma~\ref{lemma:cutfreeasr}).
  \item
    $e_n(x)=f_n(x)\cdot (e_{n-1}(g_n(x))+2g_n(x))$ for every $n\geq 1$. Indeed,
    let  $\partial(\pnone)=n\geq 1$. If $\pnone\rweal \pntwo$ in the modified level-by-level strategy, then 
    $W_{\trlalasr(\pnone)}-W_{\trlalasr(\pntwo)}\leq e_{n-1}(|\pnone|)+2|\pnone|$. This because
    we can proceed as in the case of \EAL.
    As a consequence, since $W_{\trlalasr(\pnone)}=0$ whenever $\pnone$ is cut-free 
    (again by Lemma~\ref{lemma:cutfreeasr}), we can iterate over
    the inequality $W_{\trlalasr(\pnone)}-W_{\trlalasr(\pntwo)}\leq e_{n-1}(|\pnone|)+2|\pnone|$
    obtaining $W_{\trlalasr(\pnone)}\leq f_n(|\pnone|)(e_{n-1}(g_n(|\pnone|))+g_n(|\pnone|))$ (since
    $g_n(|\pnone|)$ is a bound on the size of any reduct of $\pnone$).
\end{varitemize}
This concludes the proof.
\end{proof}}{}
By Propositions~\ref{prop:complexlal} and~\ref{prop:complexeal}, we get:
\begin{theorem}\label{th:snbound}
For every natural number $n$, there is a polynomial (resp. elementary function) 
$e_n:\N\rightarrow\N$ such that for every  term $\tmone$ 
typable in \LAL\ (resp. \EAL\ ), if $\pnone$ is a proof-net
corresponding to a type derivation of $\tmone$, then any
reduction of the sharing graph $\trlalasr(\pnone)$ (resp.
 $\trealasr(\pnone)$ ) has length 
bounded by $e_{\partial(\pnone)}(|\pnone|)$.
\end{theorem}
As an application, if $\tmone$ can be typed in \LAL\ with 
type $W \multimap \S^k W$ then there exists a polynomial
$p$ such that the application of $t$ to the term representing 
the list $w$, reduced using sharing graphs, takes at most
$p(|w|)$ steps.

\section{Soundness}\label{sect:soundness}

Suppose we are in the following situation:
\begin{displaymath}
\xymatrix{\tmone \ar[r]^>{*} \ar[d]^{\trans} & \tmtwo\\
\sgone\ar[r]^>{*} & \sgtwo}
\end{displaymath}
In words, we translated a typable term $t$ to a 
sharing graph $\sgone$, then normalised 
$\sgone$ to $\sgtwo$. We now need to define a 
read-back procedure $\mathcal{R}$ that extracts the normal form
$u$ of $t$ from $\sgtwo$. We have 
to design a variant of the readback procedures in the literature,
\textit{e.g.} \cite{GAL92,mairson02fsttcs}, since here
we are not handling a generic encoding of terms
into proof-nets but an encoding based on type derivations.

The procedure $\rbasrlm$ is
defined on sharing graphs, but does not look
at the internal structure of the graph itself;
rather, the procedure is defined as a set of 
\emph{queries} to the underlying context
semantics. \condinc{To prove $\rbasrlm$ is correct,
we will show that whenever 
$\pi:\Gamma\vdash u:A$ is a cut-free type
derivation, $\rbasrlm$ applied to the
proof-net $\pntwo$ induced by $\pi$ returns $u$.}
{To prove $\rbasrlm$ is correct,
we can suppose the input graph to come
from a cut-free proof-net $\pntwo$.}
This is enough to prove soundness. Indeed,
the context semantics of $\pntwo$ is essentially
the same as the one of $\sgtwo$:
\begin{displaymath}
\xymatrix@R=1mm@C=1mm
{
  \consem{\pnone} & \succcurlyeq & \consem{\pntwo}\\
  \| & & \\
  \consem{\sgone} & =            & \consem{\sgtwo}
}
\end{displaymath}
Observe that we could even apply the readback procedure to $G$, without
reducing $G$ to its normal form $H$. This, however, would make the
read-back process less efficient, as the whole computation
would be done by the context semantics.
\condinc{
\subsection{Structure of Normal Forms}
We here recall that for any $\lambda$-term $t$ in normal form, there are $n,m\geq 0$ such
that $t=\lambda x_1.\ldots.\lambda x_n.yt_1\ldots t_m$
where $t_1,\ldots,t_m$ are in normal form. This way, we can partition an arbitrary
normal form into head subterms. Formally, \emph{head patterns} are syntactical
objects defined by the following productions:
$$
H[\cdot]::=[\cdot]\spb\lambda x_1.\ldots.\lambda x_n.yt_1\ldots t_{i-1}H[\cdot]t_{i+1}\ldots t_{m}
$$
Given a normal form $t$, a \emph{head subterm} for $t$ is a pair 
$(H[\cdot],u)$ such that $H[\cdot]$ is a head pattern 
and $t=H[u]$. The \emph{initial head subterm} for $t$ is $([\cdot],t)$.
For example, head subterms of $t=\lambda x.y(\lambda z.zx)w$ are
\begin{eqnarray*}
\hsone_1&=&([\cdot],\lambda x.y(\lambda z.zx)w)\\
\hsone_2&=&(\lambda x.y[\cdot]w,\lambda z.zx)\\
\hsone_3&=&(\lambda x.y(\lambda z.zx)[\cdot],w)\\
\hsone_4&=&(\lambda x.y(\lambda z.z[\cdot])w,x);
\end{eqnarray*}
and the initial head subterm for $t$ is $\hsone_1$.

A function $\head$ on head subterms can be defined such that $\head(H[\cdot],u)$
returns a sequence completely describing the internal structure of $u$. In particular,
if $u=\lambda x_1.\ldots.\lambda x_n.yu_1\ldots u_m$, then $\head(H[\cdot],u)$ includes:
\begin{varitemize}
\item
  The natural number $n$.
\item
  Some binding information about the head occurrence of $y$ in $u$. More specifically,
  if the occurrence is free in $H[u]$, then $\head(H[\cdot],u)$ includes the variable $y$
  itself, while if the occurrence is bound, then $\head(H[\cdot],u)$ includes a pair
  $((J[\cdot],v),k)$ locating the binder for $y$.
\item
  The head subterms corresponding to $u_1,\ldots,u_m$.
\end{varitemize}
Formally, suppose that $t=H[J[u]]$ where
\begin{eqnarray*}
J[\cdot]&=&\lambda x_1.\ldots.\lambda x_k.yt_1\ldots t_{i-1}K[\cdot] t_{i+1}\ldots t_m;\\
u&=&\lambda z_1.\ldots.\lambda z_p.x_iu_1\ldots u_q.
\end{eqnarray*}
In this case, $\head(H[J[\cdot]],u)$ will
be the sequence
$$
(p,(H[\cdot],J[u])),i-1,(L_1[\cdot],u_1),\ldots,(L_q[\cdot],u_q))
$$ 
where for every $1\leq i\leq q$
$$
L_i[\cdot]=H[J[\lambda z_1.\ldots.\lambda z_p.x_iu_1\ldots u_{i-1}[\cdot]u_{i+1}\ldots u_q]].
$$
Now, suppose that $t=H[u]$ with
$$
u=\lambda x_1.\ldots.\lambda x_k.yt_1\ldots t_m.
$$
and the head occurrence of $y$ in $u$ is free in $t$. In this case, 
$\head(H[\cdot],u)$ will be the sequence 
$$
(k,y,(L_1[\cdot],t_1),\ldots,L_m[\cdot],t_m)).
$$
where, for every $1\leq i\leq m$,
$$
L_i[\cdot]=H[\lambda x_1.\ldots.\lambda x_p.yt_1l\ldots t_{i-1}[\cdot]t_{i+1}\ldots t_m].
$$
As an example,
\begin{eqnarray*}
\head(\hsone_1)&=&(1,y,\hsone_2,\hsone_3)\\
\head(\hsone_2)&=&(1,\hsone_2,0,\hsone_4)\\
\head(\hsone_3)&=&(0,w)\\
\head(\hsone_4)&=&(0,\hsone_1,0)
\end{eqnarray*}
For every head subterm $\hsone$, the sequence $\head(\hsone)$
is univocally determined. As a consequence, $\head$ can
be seen as a total function acting on the space of head subterms.

Let us now forget about the internal structure of head subterms
and focus on the $\head$ function. If we know which is the initial
subterm of a given term $t$, we can reconstruct the whole term $t$ by 
invoking $\head$ many times. For example, $\head(\hsone_1)$ tells
us that $t$ has $1$ abstraction, its head variable is $y$ and the
two subterms corresponds to $\hsone_2$ and $\hsone_3$. Calling again
$\head$ on $\hsone_2$, we can get some information on the first of
the two subterm: it has one abstraction, the head variable is bound
by the same abstraction and it has another subterm, corresponding to 
$\hsone_4$. And so on.

\subsection{The Read-back Procedure}\label{subsect:readbackprocedure}
The map $\headpn$ is defined as a function on pairs in the form $(\pnone,C)$
(where $\pnone$ is a proof-net and $C$ is a context for $\pnone$).
The value of $\headpn$ on $(\pnone,(p,C_1,\ldots,C_k,T))$ is defined as follows:
\begin{varitemize}
\item
  Let $n$ be the least number such that $\consem{\pnone}$ is defined 
  on $(p,C_1,\ldots,C_k,T\QQ^n)$. Suppose
  $\csor{(p,C_1,\ldots,C_k,T\QQ^n)}{(r,D_1,\ldots,D_k,S)}$.
\item
  If $S=R\PP\QQ^l\PP\QQ^m$, then return the value
  $$
  (n,(\pnone,(r,D_1,\ldots,D_k,R\PP)),l,Q_1,\ldots,Q_m)
  $$
  where for every $1\leq i\leq m$
  $$
  Q_i=(\pnone,(r,D_1,\ldots,D_k,R\PP\QQ^l\PP\QQ^i\PP))
  $$
\item
  If $S=\QQ^m$, then return the value
  $$
  (n,r,Q_1,\ldots,Q_m)
  $$
  where for every $1\leq i\leq m$
  $$
  Q_i=(\pnone,(r,D_1,\ldots,D_k,\QQ^i\PP))
  $$
\end{varitemize}
Observe that in computing $\headpn$, we use nothing but the context
semantics of $\pnone$. 
Moreover, we have the following property:
\begin{proposition}\label{prop:homomorp}
If $\pi:\Gamma\vdash t:A$ is a cut-free type derivation, then
there is a function $\hommap_\pi$ mapping head subterms of
$t$ to pairs in the form $(\pnone_\pi,C)$ (where $\pnone_\pi$ is the proof-net induced by $\pi$
and $C$ is a context for $\pnone_\pi$) satisfying the following conditions:
\begin{varenumerate}
\item\label{hom:p1}
  $\hommap_\pi([\cdot],t)=(\pnone_\pi,(p,\varepsilon,\ldots,\varepsilon))$
  where $p$ is the conclusion of $\pnone_\pi$.
\item\label{hom:p2}
  For every head subterm $\hsone$ of $t$,
  if $\head(\hsone)=(i,\hstwo,j,\hsone_1,\ldots,\hsone_n)$,
  then $\headpn(\hommap_\pi(\hsone))=(i,\hommap_\pi(\hstwo),j,\hommap_\pi(\hsone_1),\ldots,\hommap_\pi(\hsone_n))$
\item\label{hom:p3}
  For every head subterm $\hsone$ of $t$,
  if $\head(\hsone)=(i,y,\hsone_1,\ldots,\hsone_n)$,
  then $\headpn(\hommap_\pi(\hsone))=(i,p,\hommap_\pi(\hsone_1),\ldots,\hommap_\pi(\hsone_n))$
  where $p$ is the edge of $\pnone_\pi$ corresponding to the variable $y$.
\end{varenumerate}
\end{proposition}
\begin{proof}
By induction on $\pi$:
\begin{varitemize}
\item
  Suppose that:
  $$
  \pi:\infer[A]{x:A\vdash x:A}{} 
  $$
  The only head subterm of $x$ is the initial subterm $([\cdot],x)$. Moreover,
  $\head([\cdot],x)=(0,x)$. We defined $\hommap_\pi([\cdot],x)=(\pnone_\pi,(p,\varepsilon,\ldots,\varepsilon))$,
  where $p$ is the conclusion of $\pnone_\pi$. Indeed, if we apply $\headpn$ to
  $(\pnone_\pi,(p,\varepsilon,\ldots,\varepsilon))$ we obtain 
  $(0,p)$ as a result and, clearly, $p$ is the edge of $\pnone_\pi$ corresponding to variable $x$.
\item
  Suppose that:
  $$
  \pi:\infer[R_\multimap]
  {\Gamma\vdash \lambda x.\tmone:A\multimap B}
  {\rho:\Gamma,x:A\vdash \tmone:B} 
  $$  
  The head subterms of $\lambda x.\tmone$ are:
  \begin{varitemize}
  \item
    The initial head subterm $([\cdot],\lambda x.\tmone)$.
  \item
    The head subterm  $(\lambda x.H[\cdot],u)$ whenever $(H[\cdot],u)$ is a
    (non-initial) head subterm for $\tmone$.
  \end{varitemize}
  The following equalities hold:
  \begin{eqnarray*}
    \head([\cdot],\lambda x.t)&=&
    \left\{
      \begin{array}{l}
        (n+1,y,(\lambda x.J_1[\cdot],u_1),\ldots,(\lambda x.J_m[\cdot],u_m))\\
        \hspace{20pt}\mbox{if}\hspace{5pt}\head([\cdot],\tmone)=(n,y,(J_1[\cdot],u_1),\ldots,(J_m[\cdot],u_m))\\ \vspace{-6pt} \\
        (n+1,([\cdot],\lambda x.t),l+1,(\lambda x.J_1[\cdot],u_1),\ldots,(\lambda x.J_m[\cdot],u_m))\\
        \hspace{20pt}\mbox{if}\hspace{5pt}\head([\cdot],\tmone)=(n,([\cdot],\lambda x.t),l,(J_1[\cdot],u_1),\ldots,(J_m[\cdot],u_m))\\ \vspace{-6pt}\\
        (n+1,([\cdot],\lambda x.t),0,(\lambda x.J_1[\cdot],u_1),\ldots,(\lambda x.J_m[\cdot],u_m))\\
        \hspace{20pt}\mbox{if}\hspace{5pt}\head([\cdot],\tmone)=(n,x,(J_1[\cdot],u_1),\ldots,(J_m[\cdot],u_m))\\
      \end{array}
    \right.\\
    &\vspace{20pt}& \\
    \head(\lambda x.H[\cdot],u)&=& 
    \left\{
      \begin{array}{ll}
        (n,y,(\lambda x.J_1[\cdot],u_1),\ldots,(\lambda x.J_m[\cdot],u_m))\\
        \hspace{20pt}\mbox{if}\hspace{5pt}\head(H[\cdot],t)=(n,y,(J_1[\cdot],u_1),\ldots,(J_m[\cdot],u_m))\\ \vspace{-6pt} \\
        (n,(\lambda x.K[\cdot],v),l,(\lambda x.J_1[\cdot],u_1),\ldots,(\lambda x.J_m[\cdot],u_m))\\
        \hspace{20pt}\mbox{if}\hspace{5pt}\head(H[\cdot],t)=(n,(K[\cdot],v),l,(J_1[\cdot],u_1),\ldots,(J_m[\cdot],u_m))\\ \vspace{-6pt}\\
        (n,([\cdot],\lambda x.t),0,(\lambda x.J_1[\cdot],u_1),\ldots,(\lambda x.J_m[\cdot],u_m))\\
        \hspace{20pt}\mbox{if}\hspace{5pt}\head(H[\cdot],t)=(n,x,(J_1[\cdot],u_1),\ldots,(J_m[\cdot],u_m))\\
      \end{array}
    \right.
  \end{eqnarray*}
  Now, let $p$ be the main conclusion of $\pnone_\rho$, $r,r_1,\ldots,r_h$ be the premises
  of $\pnone_\rho$ (where $r$ is the premise corresponding to $x$ and $r_1,\ldots,r_h$ are the
  other ones). Analogously, let $s$ be the main conclusion of $\pnone_\pi$ and
  $q_1,\ldots,q_h$ be the premises of $\pnone_\pi$. We define $\hommap_\pi$ from
  $\hommap_\rho$ (which exists by IH) as follows:
  \begin{eqnarray*}
    \hommap_\pi([\cdot],\lambda x.\tmone)&=&(\pnone_\pi,(s,\varepsilon,\ldots,\varepsilon))\\
    \hommap_\pi(\lambda x.H[\cdot],u)&=&
    \left\{
      \begin{array}{ll}
        (\pnone_\pi,(s,C_1,\ldots,C_k,\QQ S))\\
        \hspace{20pt}\mbox{if}\hspace{5pt}\hommap_\rho(H[\cdot],u)=(\pnone_\rho,(p,C_1,\ldots,C_k,S)\\ \vspace{-6pt} \\
        (\pnone_\pi,(s,C_1,\ldots,C_k,\PP S))\\
        \hspace{20pt}\mbox{if}\hspace{5pt}\hommap_\rho(H[\cdot],u)=(\pnone_\rho,(r,C_1,\ldots,C_k,S)\\ \vspace{-6pt} \\
        (\pnone_\pi,(q_i,C_1,\ldots,C_k,S))\\
        \hspace{20pt}\mbox{if}\hspace{5pt}\hommap_\rho(H[\cdot],u)=(\pnone_\rho,(r_i,C_1,\ldots,C_k,S)\\ \vspace{-6pt} \\
      \end{array}
    \right.
  \end{eqnarray*}
  We are now able to prove conditions \ref{hom:p1} to \ref{hom:p3}. By definition, it is clear that
  condition \ref{hom:p1} is satisfied. Condition \ref{hom:p2}: suppose
  that $\head(\hsone)=(i,\hstwo,j,\hsone_1,\ldots,\hsone_n)$. We can distinguish four cases,
  depending on the shape of $\hsone$ and the way we have defined $\head$:
  \begin{varitemize}
    \item
      The following equalities hold:
      \begin{eqnarray*}
        \hsone&=&([\cdot],\lambda x.\tmone)\\
        \hstwo&=&([\cdot],\lambda x.\tmone)\\
        \forall 1\leq a\leq n.\hsone_a&=&(\lambda x.J_a[\cdot],u_a)\\
        \head([\cdot],\tmone)&=&(i-1,([\cdot],\tmone),j-1,(J_1[\cdot],u_1),\ldots,(J_m[\cdot],u_m))
      \end{eqnarray*}
      By definition, $\hommap_\pi(\hsone)=(\pnone_\pi,(s,\varepsilon,\ldots,\varepsilon))$. Moreover, by
      IH,
      $$
      \headpn(\hommap_\rho([\cdot],\tmone))=(i-1,\hommap_\rho([\cdot],\tmone),j-1,
      \hommap_\rho(J_1[\cdot],u_1),\ldots,\hommap_\rho(J_m[\cdot],u_m))
      $$
      The computation of $\headpn(\hommap_\pi(\hsone))$ is carried out very similarly 
      to the one of $\headpn(\hommap_\rho([\cdot],\tmone))$. By exploiting the way we have defined
      $\hommap_\pi$ and the way $\pnone_\pi$ is built starting from $\pnone_\rho$,
      we easily get the desired equality:
      $$
      \headpn(\hommap_\pi(\hsone))=(i,\hommap_\pi(\hstwo),j,
      \hommap_\pi(\hsone_1),\ldots,\hommap_\rho(\hsone_n))
      $$
    \item
      The following equalities hold:
      \begin{eqnarray*}
        \hsone&=&([\cdot],\lambda x.\tmone)\\
        \hstwo&=&([\cdot],\lambda x.\tmone)\\
        j&=&0\\
        \forall 1\leq a\leq n.\hsone_a&=&(\lambda x.J_a[\cdot],u_a)\\
        \head([\cdot],\tmone)&=&(i-1,x,(J_1[\cdot],u_1),\ldots,(J_m[\cdot],u_m))
      \end{eqnarray*}
      By definition, $\hommap_\pi(\hsone)=(\pnone_\pi,(s,\varepsilon,\ldots,\varepsilon))$. Moreover, by
      IH,
      $$
      \headpn(\hommap_\rho([\cdot],\tmone))=(i-1,r,
      \hommap_\rho(J_1[\cdot],u_1),\ldots,\hommap_\rho(J_m[\cdot],u_m))
      $$
      The computation of $\headpn(\hommap_\pi(\hsone))$ is carried out very similarly 
      to the one of $\headpn(\hommap_\rho([\cdot],\tmone))$. By exploiting the way we have defined
      $\hommap_\pi$ and the way $\pnone_\pi$ is built starting from $\pnone_\rho$,
      we easily get the desired equality:
      $$
      \headpn(\hommap_\pi(\hsone))=(i,\hommap_\pi(\hstwo),0,
      \hommap_\pi(\hsone_1),\ldots,\hommap_\rho(\hsone_n))
      $$
    \item
      The following equalities hold:
      \begin{eqnarray*}
        \hsone&=&(\lambda x.H[\cdot],u)\\
        \hstwo&=&(\lambda x.K[\cdot],v)\\
        \forall 1\leq a\leq n.\hsone_a&=&(\lambda x.J_a[\cdot],u_a)\\
        \head(H[\cdot],\tmone)&=&(i,(K[\cdot],v),j,(J_1[\cdot],u_1),\ldots,(J_m[\cdot],u_m))
      \end{eqnarray*}
      By IH:
      $$
      \headpn(\hommap_\rho(H[\cdot],\tmone))=(i,\hommap_\rho(K[\cdot],v),j,
      \hommap_\rho(J_1[\cdot],u_1),\ldots,\hommap_\rho(J_m[\cdot],u_m))
      $$
      The computation of $\headpn(\hommap_\pi(\hsone))$ is carried out very similarly 
      to the one of $\headpn(\hommap_\rho(H[\cdot],\tmone))$. By exploiting the way we have defined
      $\hommap_\pi$ and the way $\pnone_\pi$ is built starting from $\pnone_\rho$,
      we easily get the desired equality:
      $$
      \headpn(\hommap_\pi(\hsone))=(i,\hommap_\pi(\hstwo),j,
      \hommap_\pi(\hsone_1),\ldots,\hommap_\rho(\hsone_n))
      $$
    \item
      The following equalities hold:
      \begin{eqnarray*}
        \hsone&=&(\lambda x.H[\cdot],u)\\
        \hstwo&=&([\cdot],\lambda x.\tmone)\\
        j&=&0\\
        \forall 1\leq a\leq n.\hsone_a&=&(\lambda x.J_a[\cdot],u_a)\\
        \head(H[\cdot],\tmone)&=&(i,x,(J_1[\cdot],u_1),\ldots,(J_m[\cdot],u_m))
      \end{eqnarray*}
      By IH:
      $$
      \headpn(\hommap_\rho(H[\cdot],\tmone))=(i,r,
      \hommap_\rho(J_1[\cdot],u_1),\ldots,\hommap_\rho(J_m[\cdot],u_m))
      $$
      The computation of $\headpn(\hommap_\pi(\hsone))$ is carried out very similarly 
      to the one of $\headpn(\hommap_\rho(H[\cdot],\tmone))$. By exploiting the way we have defined
      $\hommap_\pi$ and the way $\pnone_\pi$ is built starting from $\pnone_\rho$,
      we easily get the desired equality:
      $$
      \headpn(\hommap_\pi(\hsone))=(i,\hommap_\pi(\hstwo),j,
      \hommap_\pi(\hsone_1),\ldots,\hommap_\rho(\hsone_n))
      $$
    \end{varitemize}
    Now, suppose that $\head(\hsone)=(i,y,\hsone_1,\ldots,\hsone_n)$. We can distinguish two cases,
    depending on the shape of $\hsone$ and the way we have defined $\head$:
    \begin{varitemize}
    \item
      The following equalities hold:
      \begin{eqnarray*}
        \hsone&=&([\cdot],\lambda x.\tmone)\\
        \forall 1\leq a\leq n.\hsone_a&=&(\lambda x.J_a[\cdot],u_a)\\
        \head([\cdot],\tmone)&=&(i-1,y,(J_1[\cdot],u_1),\ldots,(J_m[\cdot],u_m))
      \end{eqnarray*}
      By definition, $\hommap_\pi(\hsone)=(\pnone_\pi,(s,\varepsilon,\ldots,\varepsilon))$. Moreover, by
      IH,
      $$
      \headpn(\hommap_\rho([\cdot],\tmone))=(i-1,r_y,
      \hommap_\rho(J_1[\cdot],u_1),\ldots,\hommap_\rho(J_m[\cdot],u_m))
      $$
      The computation of $\headpn(\hommap_\pi(\hsone))$ is carried out very similarly 
      to the one of $\headpn(\hommap_\rho([\cdot],\tmone))$. By exploiting the way we have defined
      $\hommap_\pi$ and the way $\pnone_\pi$ is built starting from $\pnone_\rho$,
      we easily get the desired equality:
      $$
      \headpn(\hommap_\pi(\hsone))=(i,q_y,
      \hommap_\pi(\hsone_1),\ldots,\hommap_\rho(\hsone_n))
      $$
    \item
      The following equalities hold:
      \begin{eqnarray*}
        \hsone&=&(\lambda x.H[\cdot],u)\\
        \forall 1\leq a\leq n.\hsone_a&=&(\lambda x.J_a[\cdot],u_a)\\
        \head(H[\cdot],\tmone)&=&(i,y,(J_1[\cdot],u_1),\ldots,(J_m[\cdot],u_m))
      \end{eqnarray*}
      By IH:
      $$
      \headpn(\hommap_\rho(H[\cdot],\tmone))=(i,r_y
      \hommap_\rho(J_1[\cdot],u_1),\ldots,\hommap_\rho(J_m[\cdot],u_m))
      $$
      The computation of $\headpn(\hommap_\pi(\hsone))$ is carried out very similarly 
      to the one of $\headpn(\hommap_\rho(H[\cdot],\tmone))$. By exploiting the way we have defined
      $\hommap_\pi$ and the way $\pnone_\pi$ is built starting from $\pnone_\rho$,
      we easily get the desired equality:
      $$
      \headpn(\hommap_\pi(\hsone))=(i,y,
      \hommap_\pi(\hsone_1),\ldots,\hommap_\rho(\hsone_n))
      $$
    \end{varitemize}
  \item
  Suppose that:
  $$
  \pi:\infer[L_\multimap]
  {\Gamma,\Delta,y:A\multimap B\vdash \tmtwo\{y\tmone/x\}:C} 
  {\rho:\Gamma\vdash \tmone:A & \sigma:\Delta,x:B\vdash \tmtwo:C}
  $$  
  We can proceed as in the previous case.
  \end{varitemize}
Ths concludes the proof.
\end{proof}
A readback procedure $\rbasrlm:\grasr\rightarrow\Lambda$ is defined by
iteratively calling $\headpn$. This, by Proposition~\ref{prop:homomorp},
produces the normal form of the term we started from. Moreover, 
$\trealasr$ can be given the status of a total binary relation from
$\sleal$ to $\grasr$. We finally get:}
{

We do not present the details of the read-back procedure here (see \cite{BCDL07}).
 However, we sketch
how it can be applied to the sharing graph (in normal form)
in Figure~\ref{fig:examplesg}(b), obtaining the normal form $\tmone$
of $(\lambda x.fxx)(\lambda z.gz)$. First of all, observe that
the edge $e$ corresponds to the root of the term, while edges
$f$ and $g$ correspond to the free variables of same name.
The read-back can be performed as follows:
\begin{varitemize}
\item
  First of all, we want to know the head variable of $\tmone$. To
  do that, we query the context semantics 
  on $(e,\varepsilon,\varepsilon)$, obtaining
  $\csor{(e,\varepsilon,\varepsilon)}{(f,\varepsilon,\QQ\QQ)}$.
  This tells us that the head variable of $\tmone$ is $f$
  (since we end up in $f$) and that it is applied to 
  two arguments $\tmtwo$ and $\tmthree$ (since the rightmost stack is $\QQ\QQ$).
  The context corresponding to (this occurrence of) $f$ is
  $(f,\varepsilon,\varepsilon)$. 
\item
  Now, we can ask ourselves what is the head variable of $\tmtwo$. 
  To do that, we query the context semantics
  on $(f,\varepsilon,\PP)$. It is undefined, meaning that
  the head variable (occurrence) we are looking for lies in
  the scope of a lambda abstraction. We try
  with $(f,\varepsilon,\PP\QQ)$ obtaining
  $\csor{(f,\varepsilon,\PP\QQ)}{(g,\PP,\QQ)}$. This implies
  the head variable of $\tmtwo$ is $g$ (since we end up in $g$)
  and that it is applied to one argument, call it $\tmfour$.
  Moreover, the context corresponding to this occurrence
  of $g$ is $(g,\PP,\varepsilon)$.
\item
  The next question is: which is the head variable of 
  $\tmfour$? We proceed in the usual way, querying
  the context semantics on $(g,\PP,\PP)$ and
   $\csor{(g,\PP,\PP)}{(f,\varepsilon,\PP\PP)}$.
   Remembering that  $(f,\varepsilon,\varepsilon)$ 
 is the context corresponding to the head variable
   of $\tmone$, we can infer that the head variable
   of $\tmfour$ is bound by the first lambda-abstraction
   appearing in $\tmtwo$ (namely the first argument to
   $f$ in $\tmone$). Moreover, this variable occurrence has not
   any argument. So, we can conclude 
   $\tmone=f(\lambda z.gz)\tmthree$.
\item
  Similarly, we can query the context semantics
  to get some information on $\tmthree$ and obtain
  $\tmthree=(\lambda z.gz)$.
\end{varitemize}
This query-protocol can be generalized into a function $\rbasrlm:\grasr\rightarrow\Lambda$
whose value on a graph $\sgone$ only depends on the context semantics of $\sgone$. 
But proving $\rbasrlm$ to be correct on sharing graphs in the form
$\trealasr(\pnone)$ where $\pnone$ is cut-free is relatively easy. Thus, we get:
}
\begin{theorem}[Soundness]
The $\sleal$-graph rewriting system $(\sleal,\grasr,\rwasr,\trealasr,\rbasrlm)$ is sound.
\end{theorem}
\section{Completeness}


\begin{theorem}[Completeness]
The $\sleal$-graph rewriting system $(\sleal,\grasr,\rwasr,\trealasr,\rbasrlm)$ is complete.
\end{theorem}
\begin{proof}
 It is sufficient to observe that Theorem \ref{th:snbound}
implies that  reducing $\sgone$ will lead to a normal form $\sgtwo$.
Then it follows from the soundness result of
Section \ref{sect:soundness} that $\rbasrlm(\sgtwo)=\tmtwo$.
\end{proof}
\section{Conclusions}
We proved that Lamping's abstract algorithm is sound and complete for beta
reduction of \EAL\ (and \LAL) typable pure lambda-terms. Moreover, the 
number of graph interaction steps is shown to be bounded by the 
same kind of bounds holding for proof-net normalisation. All these results have been
established by exploiting context semantics. In particular, complexity results
have been inferred in an innovative way, being inspired by~\cite{DalLagoArxiv2005d}.

Further work includes the extension of the approach to general optimal reduction.
In the full  algorithm, however, relative bounds should take the place
of absolute bounds, since any pure lambda term can be reduced.

\textbf{Acknowledgements.} We are grateful to Harry Mairson and 
Simone Martini for stimulating discussions and advice on optimal reduction.
We also wish to thank Vincent Atassi for useful conversations.


\condinc{
\bibliographystyle{plain}
\bibliography{lameal}
}{
{\tiny\bibliographystyle{latex8}
\bibliography{lameal}}
}

\end{document}